\documentclass[conference,compsoc]{IEEEtran}

\usepackage{tikz}
\usetikzlibrary{patterns.meta,backgrounds, calc, arrows.meta, intersections, patterns, positioning, shapes.misc, fadings, through,decorations.pathreplacing,shapes}
\usepackage{amsmath}
\usepackage{enumitem}
\usepackage{dashrule}
\usepackage{amssymb}
\usepackage[normalem]{ulem} 
\usepackage[ruled,vlined]{algorithm2e}
\usepackage{listofitems}
\usepackage{xspace}
\usepackage{pgfplots}
\usepackage{makecell}
\usepackage{adjustbox}
\usepackage{minted}
\usepackage{acronym}
\usepackage{subcaption}
\usepackage{hyperref}
\usepackage{pifont}
\usepackage{amssymb}
\usepackage{censor}
\usepackage{booktabs}
\usepackage{float}
\usepackage{tcolorbox}
\usepackage{listings}
\usepackage{pifont}
\usepackage[capitalise,nameinlink,noabbrev,english]{cleveref}
\usepackage[english]{babel}
\usepackage{numprint}
\npthousandsep{,}
\makeatletter
\let\old@lstKV@SwitchCases\lstKV@SwitchCases
\def\lstKV@SwitchCases#1#2#3{}
\makeatother
\usepackage{lstlinebgrd}
\makeatletter
\let\lstKV@SwitchCases\old@lstKV@SwitchCases

\lst@Key{numbers}{none}{%
    \def\lst@PlaceNumber{\lst@linebgrd}%
    \lstKV@SwitchCases{#1}%
    {none:\\%
     left:\def\lst@PlaceNumber{\llap{\normalfont
                \lst@numberstyle{\thelstnumber}\kern\lst@numbersep}\lst@linebgrd}\\%
     right:\def\lst@PlaceNumber{\rlap{\normalfont
                \kern\linewidth \kern\lst@numbersep
                \lst@numberstyle{\thelstnumber}}\lst@linebgrd}%
    }{\PackageError{Listings}{Numbers #1 unknown}\@ehc}}
\makeatother

\setminted[c]{ %
    linenos=true,             
    numbersep=2pt,
    autogobble=true,          
    framesep=2mm,
    fontsize=\scriptsize,
    xleftmargin=2mm,
    tabsize=2,
}

\usepackage{filecontents}

\newcommand{\sysname}{TÄMU\xspace}
\newcommand{\mitee}{MiTEE\xspace}
\newcommand{\teegris}{TEEGris\xspace}
\newcommand{\optee}{OP-TEE\xspace}
\newcommand{\tsix}{T6\xspace}
\newcommand{\beanpod}{Beanpod\xspace}
\newcommand{\kinibi}{Kinibi\xspace}
\newcommand{\qsee}{QSEE\xspace}

\newcommand{\numlines}{6,000\xspace} 

\newcommand{\numgpovertimetas}{337\xspace}

\newcommand{\numdatataskinibi}{31\xspace} 
\newcommand{\numdatataskinibigp}{10\xspace} 
\newcommand{\numdatatasqsee}{41\xspace} 
\newcommand{\numdatatasqseegp}{5\xspace} 
\newcommand{\numdatatasbeanpod}{11\xspace} 
\newcommand{\numdatatasbeanpodgp}{9\xspace} 
\newcommand{\numdatatasbeanpodlibc}{11\xspace} 
\newcommand{\numdatatasbeanpodtee}{11\xspace} 
\newcommand{\numdatatasmitee}{16\xspace} 
\newcommand{\numdatatasmiteegp}{16\xspace} 
\newcommand{\numdatatasmiteelibc}{16\xspace} 
\newcommand{\numdatatasmiteetee}{12\xspace} 
\newcommand{\numdatatastsix}{6\xspace} 
\newcommand{\numdatatastsixgp}{6\xspace} 
\newcommand{\numdatatastsixlibc}{0\xspace} 
\newcommand{\numdatatastsixtee}{6\xspace} 
\newcommand{\numdatatasteegris}{34\xspace} 
\newcommand{\numdatatasteegrisgp}{30\xspace} 
\newcommand{\numdatatasteegrislibc}{34\xspace} 
\newcommand{\numdatatasteegristee}{34\xspace} 

\newcommand{\perctaswithtee}{94\%\xspace}

\newcommand{\numtasall}{67}
\newcommand{\numtasteegris}{34}
\newcommand{\numtasmitee}{16}
\newcommand{\numtasbeanpod}{11}
\newcommand{\numtastsix}{6}

\newcommand{\numapiall}{566}
\newcommand{\numgpapiall}{81}
\newcommand{\numlibcapiall}{96}
\newcommand{\numteeapiall}{389}

\newcommand{\numapiteegris}{380}
\newcommand{\numgpapiteegris}{58}
\newcommand{\numlibcapiteegris}{56}
\newcommand{\numteeapiteegris}{266}

\newcommand{\numapimitee}{151}
\newcommand{\numgpapimitee}{52}
\newcommand{\numlibcapimitee}{64}
\newcommand{\numteeapimitee}{35}

\newcommand{\numapibeanpod}{153}
\newcommand{\numgpapibeanpod}{49}
\newcommand{\numlibcapibeanpod}{26}
\newcommand{\numteeapibeanpod}{78}

\newcommand{\numapitsix}{70}
\newcommand{\numgpapitsix}{56}
\newcommand{\numlibcapitsix}{0}
\newcommand{\numteeapitsix}{14}

\newcommand{\maxbbsall}{246'997}
\newcommand{\maxbbsteegris}{152'755}
\newcommand{\maxbbsmitee}{57'227}
\newcommand{\maxbbsbeanpod}{19'643}
\newcommand{\maxbbstsix}{17'368}

\newcommand{\reachablegplibcstdall}{39}
\newcommand{\reachablegplibcstdteegris}{9}
\newcommand{\reachablegplibcstdmitee}{98}
\newcommand{\reachablegplibcstdbeanpod}{70}
\newcommand{\reachablegplibcstdtsix}{77}

\newcommand{\numgreedyninetyall}{10}
\newcommand{\numgreedyninetyteegris}{8}
\newcommand{\numgreedyninetymitee}{0}
\newcommand{\numgreedyninetybeanpod}{2}
\newcommand{\numgreedyninetytsix}{1}

\newcommand{\numemutas}{67\xspace}

\newcommand{\numtees}{four\xspace}
\newcommand{\numvulns}{17\xspace}
\newcommand{\numvulntas}{11\xspace}
\newcommand{\numvulnsdevicerepro}{11\xspace}
\newcommand{\numvulnsnotondevicerepro}{5\xspace}

\newcommand{\numndays}{six\xspace}

\newcommand{\numfuzzvuln}{12\xspace}
\newcommand{\numfuzztasteegris}{16\xspace}
\newcommand{\numfuzzcrashesteegris}{23\xspace}
\newcommand{\numfuzznotimplteegris}{23\xspace}
\newcommand{\numfuzzbugteegris}{0\xspace}
\newcommand{\numfuzzmaxbbteegris}{24'740\xspace}
\newcommand{\numfuzzbbteegris}{4643\xspace}
\newcommand{\numfuzztasmitee}{10\xspace} 
\newcommand{\numfuzzcrashesmitee}{98\xspace} 
\newcommand{\numfuzznotimplmitee}{14\xspace} 
\newcommand{\numfuzzbugmitee}{84\xspace} 
\newcommand{\numfuzzmaxbbmitee}{36'128\xspace} 
\newcommand{\numfuzzbbmitee}{10'820\xspace} 
\newcommand{\numfuzztasbeanpod}{2\xspace} 
\newcommand{\numfuzzcrashesbeanpod}{2\xspace} 
\newcommand{\numfuzznotimplbeanpod}{2\xspace} 
\newcommand{\numfuzzbugbeanpod}{0\xspace} 
\newcommand{\numfuzzmaxbbbeanpod}{2485\xspace} 
\newcommand{\numfuzzbbbeanpod}{325\xspace} 
\newcommand{\numfuzztastsix}{2\xspace} 
\newcommand{\numfuzzcrashestsix}{31\xspace} 
\newcommand{\numfuzznotimpltsix}{3\xspace} 
\newcommand{\numfuzzbugtsix}{28\xspace} 
\newcommand{\numfuzzmaxbbtsix}{10'147\xspace} 
\newcommand{\numfuzzbbtsix}{848\xspace} 
\newcommand{\numfuzztas}{30\xspace} 
\newcommand{\numfuzzcrashes}{154\xspace} 
\newcommand{\numfuzznotimpl}{42\xspace} 
\newcommand{\numfuzzbug}{112\xspace} 
\newcommand{\numfuzzmaxbb}{73'500\xspace} 
\newcommand{\numfuzzbb}{16'473\xspace}

\begin{document}

\date{}

\title{\Large \bf \sysname: Emulating Trusted Applications at the (GlobalPlatform)-API Layer} 

\newcommand{\asterisk}{1}
\newcommand{\ubc}{$\dagger$}
\newcommand{\sci}{{\tiny$\infty$}}
\newcommand{\serenitix}{{$\ddagger$}}
\newcommand{\manchester}{\tiny$\nabla$}

\author{\IEEEauthorblockN{Philipp Mao\textsuperscript{\ubc}, Li Shi\textsuperscript{\serenitix}, Marcel Busch\textsuperscript{\ubc}, Mathias Payer\textsuperscript{\ubc}}
\IEEEauthorblockA{\textit{\textsuperscript{\ubc}EPFL, \textsuperscript{\serenitix}DarkNavy}}}


\maketitle

\begin{abstract}

Mobile devices rely on Trusted Execution Environments (TEEs) to execute
security-critical code and protect sensitive assets. This
security-critical code is modularized in components known as
Trusted Applications (TAs). Vulnerabilities in TAs can compromise the TEE and, thus, the entire system. 
However, the closed-source nature and fragmentation of mobile TEEs severely hinder dynamic
analysis of TAs, limiting testing efforts to mostly static analyses.

This paper presents \sysname, a rehosting platform enabling dynamic analysis of
TAs, specifically fuzzing and debugging, by interposing their execution at
the API layer. To scale to many TAs across different TEEs,
\sysname leverages the standardization of TEE APIs, driven by the
GlobalPlatform specifications. For the remaining TEE-specific APIs not shared
across different TEEs, \sysname introduces the notion of \emph{greedy
high-level emulation}, a technique that allows prioritizing manual rehosting
efforts based on the potential coverage gain during fuzzing. We implement
\sysname and use it to emulate \numemutas TAs across \numtees TEEs.  Our
fuzzing campaigns yielded \numvulns zero-day vulnerabilities across
\numvulntas TAs. These results indicate a deficit of dynamic analysis
capabilities across the TEE ecosystem, where not even vendors \emph{with
source code} unlocked these capabilities for themselves. \sysname promises to
close this gap by bringing effective and practical dynamic analysis to the
mobile TEE domain.

\end{abstract}

\section{Introduction}\label{sec:introduction}

Trusted Applications (TAs) run in the user space on top of the 
Trusted Execution Environment (TEE) Operating System (TOS). Each TA encapsulates 
security-critical functionality such as fingerprint verification, digital rights 
management, or payment services, which is exposed to the untrusted normal world.
Because TAs communicate with the normal world and process untrusted input, their 
security is critical. Bugs in TAs have repeatedly been used to 
achieve full-device compromise~\cite{beniaini2017trustissues,berard2018kinibi,busch2020finding,rollback,globalconfusion,Komaromy2021UnboxYourPhone,Li2024DiveIntoAndroidTA,AdamskiGuilbonPeterlin2019SamsungTrustZone,Makkaveev2019TrustZoneFuzzing,horta,AdamTC}.  

Dynamic analysis, such as fuzzing or debugging, helps by proactively 
analyzing TAs to discover bugs. However, due to TAs running in the TEE, 
dynamic analysis on-device, with introspection capabilities that go beyond 
the TEE log output, is not possible as only correctly signed code 
can run in the TEE. The alternative to unlocking dynamic analysis capabilities is emulation. 
However, emulating TAs is challenging due to the heterogeneity of real-world TEE implementations. 
TEEs are completely proprietary with vastly different TOSes and runtimes. 
For example, Xiaomi's \mitee uses a Fuchsia-based microkernel, while \teegris by Samsung 
is a monolithic kernel with some POSIX support as well as custom system calls. These TOSes  
have drastically different system call interfaces.  
Emulating a TA requires supporting the underlying interactions with the TOS. Due to the 
heterogeneity of TOSes, emulation effort spent on one TOS does not translate to support for others.

Samsung's PartEmu~\cite{partemu} uses full-system emulation to boot and run the entire TOS 
and TAs. While this approach transparently handles TA-to-TOS interactions,
 the prototype has neither been open-sourced nor has it been reproduced due to the in-house 
TOS details needed for this approach. Researchers have instead turned to 
user mode emulation of TAs~\cite{lightemu,fan2025qsee,qseecyberes,Thalium2023PivotingSecureWorld,Komaromy2021UnboxYourPhone,AdamskiGuilbonPeterlin2019SamsungTrustZone,Li2024DiveIntoAndroidTA}, 
implementing the specific TOS system 
calls or hooking API-level interactions. While these approaches have proven to 
be practical, they often focus on a single TEE or use an ad-hoc 
combination of API-level and system call hooking.

\sysname presents a principled approach to TA emulation, applying high-level emulation (HLE) at the 
API level to TAs, a technique we refer to as API-level interception. \sysname first leverages the de-facto 
standardization by means of the GlobalPlatform (GP) APIs.
Many TAs are not developed by the TEE vendor but by a third party (such as Google 
for the Widevine TA) and thus need to be deployable on various TEEs.
Due to the fragmentation of TEEs, the GP specifications were
introduced to enable interoperability for TAs between TEEs.
The GP specifications define entrypoint functions in the TA exposed to the normal world and a set of APIs that  
the TEE is expected to implement for a TA, including memory management, cryptography, and storage. 
The goal of the GP API is to provide an abstract layer for the concrete 
underlying TEE implementation. \sysname interposes TAs at the API level and leverages
the standardization of TA APIs to scale its emulation approach to 
diverse TEEs.

We conduct a study of the TEEs deployed on modern Android smartphones,  
quantify their level of API standardization, and discuss 
the feasibility of supporting these TAs on \sysname. 
We find that although the GP API is widely adopted, TAs also make 
use of standard libc APIs and proprietary TEE-specific APIs.
Adding support to \sysname for libc APIs 
is straightforward as these APIs are publicly documented.
Unfortunately, a majority of 
TAs still makes use of proprietary TEE-specific functionality. 
\perctaswithtee of all TAs in our dataset use at least one TEE-specific API. 
Faithfully emulating TEE-specific APIs requires significant manual effort: 
reverse engineering of the API's functionality, and then implementing 
that functionality in an emulator shim.

To handle the manual effort incurred by supporting TEE-specific APIs, we introduce 
greedy HLE, a technique that leverages static analysis to calculate the 
potentially reachable code as a function of the set of implemented APIs.
This approach allows us to prioritize TEE-specific APIs based on the expected coverage yield.
Leveraging greedy HLE, we first demonstrate that with just the GP and libc APIs, \reachablegplibcstdall\%
of code can be reached.
For each TEE, there are between one to eight TEE-specific APIs, which, after being implemented 
in \sysname, enable execution of up to 90\% of basic blocks.
We further demonstrate how 70\% of the most impactful TEE-specific APIs (in terms of newly reachable 
basic blocks) can realistically be replaced by a GP or libc API, reducing the implementation effort 
and paving the way for further standardization.

We implement \sysname and show that its API interception 
approach is feasible by correctly running~\numemutas TAs across \numtees TEEs.  
We demonstrate the fidelity and usefulness of \sysname by reproducing 
n-day vulnerabilities. We further fuzz \numfuzztas TAs and discover \numvulns 0-day vulnerabilities. 
We have responsibly disclosed all the vulnerabilities to the affected vendors.
\sysname is open source and available at \url{https://github.com/HexHive/taemu}.
In summary, we make the following contributions:

\begin{itemize}

     \item We identify, measure, and leverage the adoption of the GP APIs 
    as a means to bring cost-effective HLE to the TA ecosystem. Our 
    insight reveals the economies of scale unlocked by the increasing adoption of GP APIs.

    \item We propose a rehosting methodology (greedy HLE) for proprietary TA APIs based on data-driven 
    decisions to prioritize rehosting efforts.

    \item We design and implement \sysname, an open-source implementation of our API interception 
    approach, able to run \numemutas TAs with an integrated fuzzing component that has discovered 
    \numvulns 0-day vulnerabilities.

\end{itemize}

\section{High-Level Emulation and GlobalPlatform API Adoption}\label{sec:background}

The core intuition of \sysname results from the observation that the TA ecosystem is 
adopting a common API, the GP TEE API, that unlocks unprecedented 
economies of scale, thereby mitigating the major challenges faced by previous rehosting works. 
In this section, we discuss the prior attempts and pitfalls to leverage HLE in other domains, 
provide the background on the GP APIs, and 
empirically validate our industry adoption claims regarding the APIs in question.

\subsection{Prior Art}
HLE is an approach where, instead of faithfully
emulating every hardware peripheral at the register level, the emulator
intercepts calls to well-defined software libraries or hardware abstraction
layers (HALs) and re-implements their functionality on the host. This allows
firmware to run without requiring a fully precise peripheral model and can greatly
accelerate re-hosting. For example, when emulating a network card, instead of 
implementing all hardware intricacies, HLE would interpose the emulation at the 
send/receive packet API. However, in the deeply embedded world, where this idea 
was proposed initially by HALucinator~\cite{halucinator}, it has seen little adoption because 
the HAL landscape is highly fragmented: each silicon
vendor, and often each chip family, ships its own HAL with different APIs,
naming conventions, and binary layouts. Covering this diversity requires
building and maintaining a large set of HLE shims for every variant,
undermining the scalability and portability that make HLE attractive in the
first place. HALucinator explores this space but is limited by the 
lack of a unified HAL standard, limiting its scope to a few targets 
where HALs are inferred manually, resulting in limited scalability and therefore 
little adoption. 

\begin{tcolorbox}
    HLE emulation is feasible in scenarios where a 
    well-defined, unified abstraction interface exists.
\end{tcolorbox}

\subsection{GlobalPlatform APIs}\label{sec:gpapis}

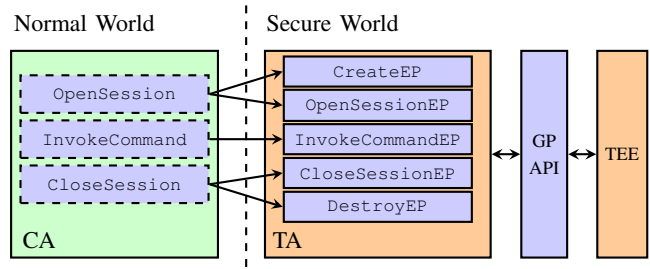
\begin{figure}[]
  \centering
  \centering
\begin{tikzpicture}
    \small
    \coordinate (normal) at (-2.5,0);
    \coordinate (secure) at (1,0);
        
\begin{scope}[shift=(normal)]
    \coordinate (clientapi) at (0,0.2);
    \node[draw, fill=green!20, minimum width=2.75cm, minimum height=2.75cm, thick] (ca) at (0,0) {};
    \begin{scope}[shift=(clientapi)]
        \node[draw, dashed, fill=blue!20, minimum width=2.5cm, minimum height=0.5cm, thick] (copen) at (0,0.6) {\scriptsize \texttt{OpenSession}};
        \node[draw, dashed, fill=blue!20, minimum width=2.5cm, minimum height=0.5cm, thick] (cinvok) at (0,0) {\scriptsize \texttt{InvokeCommand}};
        \node[draw, dashed, fill=blue!20, minimum width=2.5cm, minimum height=0.5cm, thick] (cclose) at (0,-0.6) {\scriptsize \texttt{CloseSession}};
    \end{scope}
    \node[align=center] at ($(ca.south) + (-1.0cm,0.25cm)$) {CA};
\end{scope}

\begin{scope}[shift=(secure)]
    \coordinate (tafw) at (0,0);
    \node[draw, fill=orange!40, minimum width=3cm, minimum height=2.75cm, thick] (ta) at (0,0) {};
    \begin{scope}[shift=(tafw)]
        \node[draw, fill=blue!20, minimum width=2.5cm, minimum height=0.4cm, thick] (create) at (0,1.1) {\scriptsize \texttt{CreateEP}};
        \node[draw, fill=blue!20, minimum width=2.5cm, minimum height=0.4cm, thick] (open) at (0,0.65) {\scriptsize \texttt{OpenSessionEP}};
        \node[draw, fill=blue!20, minimum width=2.5cm, minimum height=0.4cm, thick] (invoke) at (0,0.2) {\scriptsize \texttt{InvokeCommandEP}};
        \node[draw, fill=blue!20, minimum width=2.5cm, minimum height=0.4cm, thick] (close) at (0,-0.25) {\scriptsize \texttt{CloseSessionEP}};
        \node[draw, fill=blue!20, minimum width=2.5cm, minimum height=0.4cm, thick] (destroy) at (0,-0.7) {\scriptsize \texttt{DestroyEP}};
    \end{scope}
    
    \node[draw, fill=blue!20, align=center, minimum width=0.5cm, minimum height=2.75cm, thick] (gpapi) at (2.2125,0) {\scriptsize GP\\\scriptsize API};
    
    \node[draw, fill=orange!40, minimum width=0.5cm, minimum height=2.75cm, thick] (tee) at (3.25,0) {\scriptsize TEE};
    \node[align=center] at ($(ta.south) + (-1.2cm,0.25cm)$) {TA};

\end{scope}

\draw[-, thick, dashed] (-0.75,-1.5) -- (-0.75,2);
\node at (-2.9,1.75) {Normal World};    
\node at (0.4,1.75) {Secure World};    

\draw[->, thick, >=stealth] (copen.east) -- (create.west) node[midway, above, sloped] {};
\draw[->, thick, >=stealth] (copen.east) -- (open.west) node[midway, above, sloped] {};
\draw[->, thick, >=stealth] (cinvok.east) -- (invoke.west) node[midway, above, sloped] {};
\draw[->, thick, >=stealth] (cclose.east) -- (close.west) node[midway, above, sloped] {};
\draw[->, thick, >=stealth] (cclose.east) -- (destroy.west) node[midway, above, sloped] {};
\draw[<->, thick, >=stealth] ($(gpapi.west) + (0,0.)$) -- ($(ta.east) + (0,0)$) node[midway, above, sloped] {};
\draw[<->, thick, >=stealth] ($(gpapi.east) + (0,0.)$) -- ($(tee.west) + (0,0)$) node[midway, above, sloped] {};

\end{tikzpicture}
  \caption{
    An overview of how GP facilitates communication between the normal world and TA, 
    and between TA and TEE (The TEE includes the user space runtime environment and the 
    TOS kernel). The dashed blue boxes are part of the GP client API, while the solid blue 
    boxes are part of the GP Internal Core API.
    }
  \label{fig:gpover}
\end{figure}

Unlike embedded HALs, TEEs have begun to converge around a standardized abstraction layer — the GlobalPlatform (GP) API.

The TEE landscape has historically been fragmented,
with each vendor implementing its own proprietary APIs and runtime services.
TAs built for one TEE often require significant modifications to run on
another, making cross-platform deployment expensive and error-prone. This
fragmentation is further amplified by device manufacturers shipping products
with diverse system-on-chips and incompatible TEE stacks, forcing developers to
maintain multiple codebases and weakening the portability of security-critical
components.

To address this challenge, GP proposes specifications, which
define a unified interface for both the normal  and the secure world. 
GP provides a consistent programming model that abstracts vendor-specific differences. This
standardization enables developers to write TAs once and deploy them across any
GP-compliant TEE with minimal adaptation, reducing engineering overhead and
fostering a more interoperable and secure ecosystem.

The GP specification is split into two complementary parts. The Client
API~\cite{gpclientapi} defines how normal-world applications communicate with
TAs through standardized session management and parameter passing. In contrast,
the TEE Internal Core API~\cite{gpintapi} specifies the secure-world
environment in which TAs run, including entrypoints, which handle requests from the normal world, 
as well as core services such as memory management, cryptography, secure storage, and
inter-TA communication. 
Together, these two APIs bridge the gap between heterogeneous hardware and portable, 
security-critical software. \autoref{fig:gpover} gives an overview of how the GP API 
bridges the normal and the secure world.

\begin{figure}[]
    \input{listings/ta.tex}                                                                                                                     \captionof{lstlisting}{
        An example TA's GP entrypoint function (\texttt{TA\_InvokeCommandEntryPoint}). 
        The TA encrypts data with a TEE-internal key and returns the ciphertext to the normal world. 
        Lines marked in \colorbox{blue!40}{blue} are instances of the TA calling GP APIs. Lines marked 
        in \colorbox{teal!40}{green} contain a call to a libc API. 
        Lines in \colorbox{red!40}{red} are calls to TEE-specific APIs.
    }
    \label{lst:ta}
\end{figure}

By adhering to the GP specification, a TA developer can deploy their TA on different 
GP-compliant TEEs without heavily modifying the TA for each TEE.
With the exception of Google's Trusty, all TEEs deployed on modern Android are at 
least partially GP-compliant. While the GP APIs are commonly used, many TAs also 
rely on libc standard API functions. Furthermore, TAs may also include TEE-specific 
APIs, functions not defined in either the GP or libc specification. 
Listing~\ref{lst:ta} shows the source code of an example TA's 
\texttt{TA\_InvokeCommandEntryPoint} function (a GP entrypoint function handling requests 
from the normal world). 
The TA uses the GP API to allocate memory and the GP cryptographic API to encrypt the user input. It also uses the libc 
API to make a copy of the input buffer. Furthermore, it uses TEE-specific APIs 
to retrieve the encryption key and for logging. When porting this TA from one TEE to another TEE 
supporting GP and libc APIs, the developer only needs to adjust the two TEE-specific API invocations.

\begin{tcolorbox}
    In the TA space, GP and libc APIs define a clear abstraction layer, 
    whose adoption is driven by the need for standardization in the TEE market.
\end{tcolorbox}

\subsection{GP API Adoption Measurement}\label{sec:gpadoption}

\begin{table}[]
    \footnotesize
    \centering
    \begin{tabular}{lll}

        TEE&
        SOCs &
        \makecell{OEMs}\\
        \toprule
        
        \teegris & Exynos & Samsung\\
        \mitee & MediaTek & Xiaomi \\
        \beanpod & MediaTek & Xiaomi \\
        \tsix & MediaTek & various \\
        \qsee & QualComm & various\\
        \kinibi & \makecell{Exynos, MediaTek} & Samsung + various \\
        TrustedCore & HiSilicon & Huawei \\

        \bottomrule
    \end{tabular}
    \caption{
        An overview of the TEEs considered in the GP API adoption measurement study.
    }
    \label{tab:tees}
\end{table}

Before investing significant engineering effort into rehosting TAs for dynamic
analysis and fuzzing, it is essential to understand how much code actually
shares a set of common APIs. Rehosting is costly: it requires reverse
engineering, emulator extension, and often custom peripheral modeling. If the
ecosystem is highly fragmented and lacks shared interfaces, each TEE would
require its own set of high-level emulation (HLE) shims, making large-scale
dynamic testing impractical. Conversely, if most TAs rely on a common standard,
rehosting becomes far more scalable, as one emulator backend can support a wide
range of devices.

\begin{figure}[t]
    \definecolor{beanpodplt}{rgb}{0.12156862745098039,0.4666666666666667,0.7058823529411765}
    \definecolor{kinibiplt}{rgb}{0.6823529411764706,0.7803921568627451,0.9098039215686274}
    \definecolor{miteeplt}{rgb}{1.0,0.4980392156862745,0.054901960784313725}
    \definecolor{qseeplt}{rgb}{1.0,0.7333333333333333,0.47058823529411764}
    \definecolor{t6plt}{rgb}{0.17254901960784313,0.6274509803921569,0.17254901960784313}
    \definecolor{teegrisplt}{rgb}{0.596078431372549,0.8745098039215686,0.5411764705882353}
    \definecolor{trustedcoreplt}{rgb}{0.8392156862745098,0.15294117647058825,0.1568627450980392}
    \begin{tikzpicture}
    
        \node[] at (-1.2,1.7) [] {\footnotesize{80}};
        \node[] at (-1.2,0.85) [] {\footnotesize{40}};
        \node[] at (-1.2,0) [] {\footnotesize{0}};
    
        \node[] at (-1.2,-0.85) [] {\footnotesize{40}};
        \node[] at (-1.2,-1.7) [] {\footnotesize{80}};

        \node[anchor=center] (all) at (1,0) {
            \includegraphics[width=4cm]{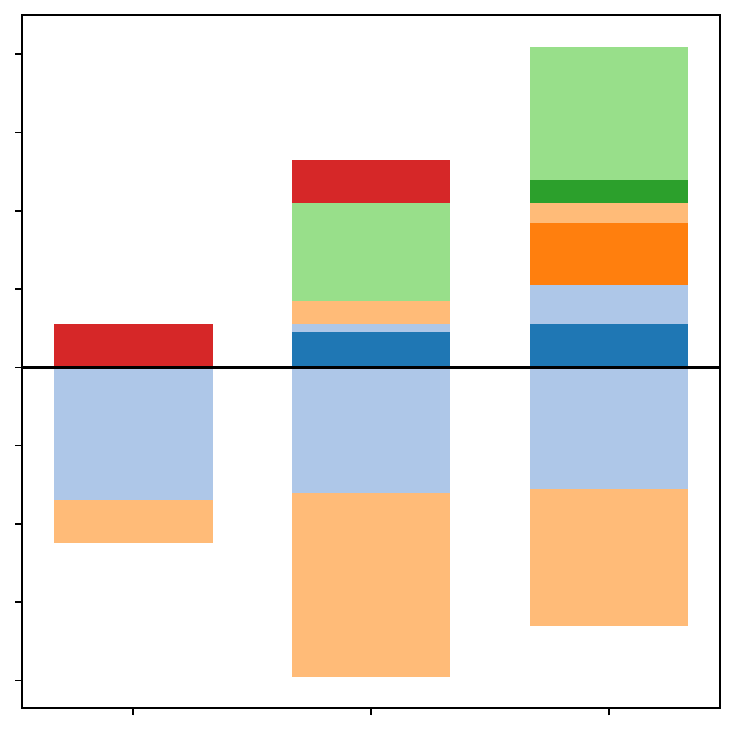}
        };

        \node[rotate=90, anchor=west] at (all.west) [xshift=-0.5cm, yshift=0.5cm] {\footnotesize{\# TA}};

        \node at (all.west) [yshift=1.5cm,xshift=0.8cm] {GP};
        \node at (all.west) [yshift=-1.5cm,xshift=1cm] {Custom};
        \node at (all.south) [yshift=-0cm,xshift=-1.3cm] {\footnotesize 2015};
        \node at (all.south) [yshift=-0cm,xshift=0cm] {\footnotesize 2020};
        \node at (all.south) [yshift=-0cm,xshift=1.3cm] {\footnotesize 2025};

        \node[anchor=north west] (legend) at ([xshift=0.5cm]all.north east) {};
        \foreach \color/\name [count=\i] in {
            beanpodplt/\beanpod,
            kinibiplt/\kinibi,
            miteeplt/\mitee,
            qseeplt/\qsee,
            t6plt/\tsix,
            teegrisplt/\teegris,
            trustedcoreplt/TrustedCore
        } {
            \node[draw, fill=\color, minimum size=4pt, inner sep=0pt, anchor=north west] 
            at ([yshift=-\i*0.5cm]legend.north west) {};
        \node[anchor=west] 
            at ([xshift=0.25cm,yshift=-\i*0.5cm-0.1cm]legend.north west) {\footnotesize \name};
        }

    \end{tikzpicture}
    \caption{
        The distribution of TAs using the GP API compared to TAs without across three time snapshots. 
    }
    \label{fig:gpovertime}
\end{figure}

We perform an empirical study combining static inspection of TA binaries with
historical snapshots to map adoption trends over time. We download firmware for various system on chip and phone vendors 
from three points in time (2015, 2020, and 2025). We extract the TA binaries from the firmware and semi-automatically  
analyze them with a decompiler. To determine whether a TA makes use of GP APIs, we look for the GP entrypoint 
functions. To search for the GP entrypoint functions, we first use a script 
that marks TAs for further manual analysis, if at least one potential function (matching function signature) 
is found for each GP entrypoint function. We then manually analyze the marked TAs.
We mark TAs as using the GP API if we confirm the existence of the GP entrypoint functions.
This allows us to identify when major vendors began integrating GP APIs and how consistently
developers now depend on them.
We analyze 
\numgpovertimetas{} TAs covering seven commercially used and proprietary TEE implementations. 
To the best of our knowledge, these are 
\emph{all} the TEEs currently deployed on Android phones, excluding Huawei's iTrustee, 
whose TAs are encrypted, and Google's Trusty, which is only used on Pixel devices 
and does not reuse the GP specifications. See \autoref{tab:tees} for additional information 
on the considered TEEs.
\autoref{fig:gpovertime} 
shows the trend: only 19\% of TAs relied on the GP API in 2015, 
adoption reached 40\% in 2020, and exceeded 55\% in 2025. All TEEs introduced after 
2020 make use of the GP API. 
The number of TAs not supporting GP has stagnated and is 
confined to \kinibi and \qsee with the remaining five TEEs supporting the GP API.
Even for \kinibi and \qsee, new TAs are written to be GP-compliant.

\begin{table}[]
    \footnotesize
    \centering
    \begin{tabular}{rrr rrr}

        TEE&
        \textbf{\# TAs} &
        \makecell{\# TAs\\GP-EP} &
        \makecell{\# TAs\\GP API}  &
        \makecell{\# TAs\\libc API} &
        \makecell{\# TAs\\TEE API} \\

        \cmidrule(r){1-1} \cmidrule(r){2-3} \cmidrule(r){4-6}

        \teegris & \textbf{\numdatatasteegris} & \numdatatasteegris & \numdatatasteegrisgp & \numdatatasteegrislibc & \numdatatasteegristee\\
        \mitee & \textbf{\numdatatasmitee} & \numdatatasmiteegp & \numdatatasmiteegp & \numdatatasmiteelibc & \numdatatasmiteetee \\
        \beanpod & \textbf{\numdatatasbeanpod} & \numdatatasbeanpod & \numdatatasbeanpodgp & \numdatatasbeanpodlibc & \numdatatasbeanpodtee \\
        \tsix & \textbf{\numdatatastsix} & \numdatatastsixgp & \numdatatastsixgp & \numdatatastsixlibc & \numdatatastsixtee \\

        \cmidrule(r){1-1} \cmidrule(r){2-3} \cmidrule(r){4-6}

        \qsee & \textbf{\numdatatasqsee} & \numdatatasqseegp & \numdatatasqseegp & 0 & \numdatatasqseegp\\
        \kinibi & \textbf{\numdatataskinibi} & \numdatataskinibigp & 0 & 0 & \numdatataskinibigp\\

        \bottomrule
    \end{tabular}
    \caption{Our dataset of TAs from 2025 along with the types of API used by TAs. For TAs that expose the 
    GP entrypoints (EP), we further analyze if the TA uses the GP APIs, libc APIs, or TEE-specific APIs.}
    \label{tab:gpcomp}
\end{table}

\begin{figure}[t]
    \centering
    \includegraphics[width=\columnwidth, keepaspectratio]{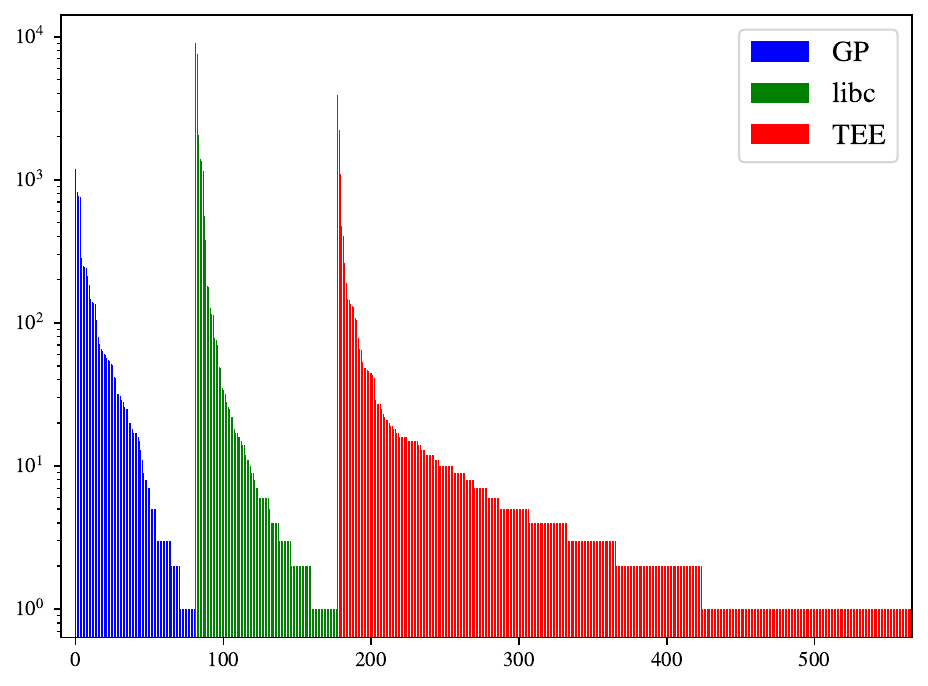}
    \caption{
    The frequency graph of the types of APIs used across our dataset of GP-compliant TAs. The y-axis denotes 
    the number of invocations of a given API function in log scale. 
    } 
    \label{fig:freq}
\end{figure}

As discussed in~\autoref{sec:gpapis}, a TA using the GP APIs may also rely on 
libc and TEE-specific APIs. We analyze the 82 GP-compliant TAs from 2025 and whether these TAs 
make use of libc or TEE-specific APIs. The results are shown in \autoref{tab:gpcomp}.
TAs from \teegris, \mitee, and \beanpod make heavy use of functions from the libc standard library.
Supporting libc APIs for HLE is similar to supporting GP APIs, since their functionality is publicly 
documented. 
Thus, we treat libc APIs the same as GP APIs, i.e., standard APIs, whose HLE emulation shim scales to various TEEs.

\begin{tcolorbox}
    The empirically demonstrated convergence towards standardization indicates that the
    TEE ecosystem is now a prime candidate for HLE-based dynamic analysis, where
    one GP/libc-aware rehosting framework could scale across most TEEs.
\end{tcolorbox}

Critically, we find that \perctaswithtee of TAs use at least one 
TEE-specific function. \autoref{fig:freq} shows the frequency graphs of the APIs used by the 82 TAs. Each point on 
the x-axis is an API function (either GP, libc, or TEE-specific). The y-axis denotes in log scale the number of times the API is called across all TAs.
All three categories contain a few high-frequency functions. However, there are more TEE-specific APIs both in terms of the number of 
unique functions and invocations.

\begin{tcolorbox}
    TEE-specific APIs are still widely used and account for a substantial 
    portion of TA API invocations. Therefore, an effective rehosting approach must 
    address TEE-specific APIs.
\end{tcolorbox}

\section{\sysname}\label{sec:desgin} 

\begin{figure*}[th!]
  \centering
  \input{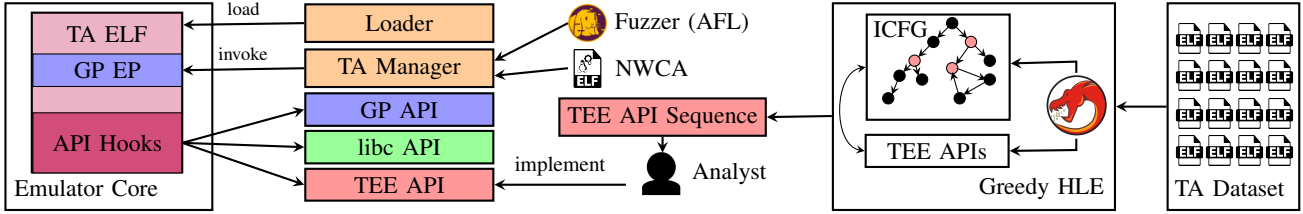}
  \caption{Overview of \sysname's design. 
    }
  \label{fig:design}
\end{figure*}

\autoref{fig:design} illustrates \sysname's design.
The left side shows \sysname's rehosting platform, which allows 
the execution of GP-compliant TAs on commodity hardware through 
API-level interception. We explain how the components on the left side work to enable 
API-level interception in \autoref{sec:apidesign}.
The right side shows greedy HLE, a technique to prioritize TEE-specific APIs based 
on the potential coverage gain. In \autoref{sec:greedyhle}, we discuss this technique 
and how it enables \sysname to scale in the face of TEE-specific APIs.
The middle of the figure shows how an analyst interacts with \sysname. The analyst extends 
\sysname's TEE-specific APIs using the TEE APIs obtained with greedy HLE. An analyst interacts with the emulated TA either via commodity fuzzers (AFL) or normal world client applications (NWCA).

\subsection{API-Level Interception}\label{sec:apidesign}

At the core of \sysname's design is the insight that the TA ecosystem is moving 
towards standardization in terms of the APIs leveraged by TAs. \sysname's 
rehosting platform is built around a virtual TEE that is able to load, invoke, and provide 
the required standard GP and libc APIs to the emulated TA. The TA itself runs in 
the emulator core, which implements the processor emulation---the
semantically correct execution of instructions and the corresponding CPU register
and memory updates. The emulator core enables \sysname to run TAs from any architecture 
(primarily arm32 or arm64) on a commodity x86 processor.

\sysname's loader component takes care of correctly mapping the TA into the emulator core. 
All of \sysname's supported TEEs ship their TAs as ELF files, allowing \sysname's 
loader to reuse existing ELF loading functionality. However, there is no inherent limitation to 
extend the loader to support TEE-specific executable formats, such as \kinibi's proprietary 
MCLF format.

To enable API-level interception, \sysname instruments the TA's API invocations such that whenever the 
TA invokes an API function, execution is redirected to \sysname's virtual TEE. From there, 
the corresponding HLE API implementation in \sysname is called. Afterwards, execution is resumed in the TA 
at the expected return address.
API invocations are instrumented by installing hooks in the emulator core, which 
trigger whenever the instruction pointer is equal to the address of an API invocation.
The instruction pointer value is used to infer which API the TA was calling. 

While these are standard techniques for HLE, \sysname's main contribution stems from 
the insight that, thanks to the TEE ecosystem's standardization, heterogeneous TEE interactions 
go through standard GP or libc APIs. \sysname's virtual TEE hooks and implements both the GP 
and libc APIs. Implementing these APIs correctly is feasible due to both GP and libc 
providing specifications for their APIs. Furthermore, these APIs behave the same regardless  
of the underlying TEE and implementing them is a one-time effort that scales to any TEE. 
We differentiate between the following categories of standard APIs:

\textbf{Stateless APIs.} Stateless APIs behave the same irrespective of any prior API invocations 
by the TA. Examples of such APIs are libc's \texttt{memcmp} or GP's \texttt{TEE\_MemCompare}. 

\textbf{Stateful APIs.} These APIs behave differently depending on prior API invocations.
\sysname tracks such internal state to accurately emulate these APIs. Examples of such APIs 
are memory allocation APIs such as \texttt{malloc}/\texttt{TEE\_Malloc} and GP APIs related 
to storage and cryptography. 

\textbf{GP Entrypoints.} Unlike the previously described APIs, the GP entrypoint APIs in \sysname 
are not invoked by the TA itself but by \sysname's TA manager component. Invocation of an entrypoint 
triggers the emulation of the TA until the entrypoint function returns. The TA manager calls these 
entrypoint functions in response to interactions from external programs communicating with the emulated TA.
The TA manager supports two modes of interaction: \emph{interactive} and \emph{fuzzing}. 
In interactive mode, a NWCA can communicate with the emulated TA 
just as it would on a real device. In fuzzing mode, the TA manager acts as a harness for a fuzzer, 
deserializing the fuzzing input into the TA's input buffers and invoking the TA's command-handling entrypoint API.

As discussed in \autoref{sec:gpadoption}, TEE-specific APIs make up a significant 
amount of the API invocations in our dataset. While the technique to intercept TEE-specific 
APIs is the same as described above, there is no specification for these TEE-specific APIs, and thus 
implementing them requires manual effort. We discuss how \sysname handles TEE-specific APIs 
leveraging greedy HLE in the following subsection. 

\subsection{Greedy HLE}\label{sec:greedyhle}

\begin{table*}[h]
    \small
    \centering
    \begin{tabular}{lrrrrrrrr}
        TEE&
        \# TAs & 
        \# APIs &
        \makecell{\# GP APIs} &
        \makecell{\# libc APIs} &
        \makecell{\# TEE APIs} &
        \makecell{\# max. BBs}&
        \makecell{std-cov.*}&
        \makecell{\# greedy 90\%\dag}\\
        \midrule
        \teegris & \numtasteegris & \numapiteegris & \numgpapiteegris & \numlibcapiteegris & \numteeapiteegris & \maxbbsteegris & \reachablegplibcstdteegris{}\%& \numgreedyninetyteegris\\
        \mitee & \numtasmitee & \numapimitee & \numgpapimitee & \numlibcapimitee & \numteeapimitee & \maxbbsmitee & \reachablegplibcstdmitee{}\% & \numgreedyninetymitee\\
        \beanpod & \numtasbeanpod & \numapibeanpod & \numgpapibeanpod & \numlibcapibeanpod & \numteeapibeanpod & \maxbbsbeanpod & \reachablegplibcstdbeanpod{}\%& \numgreedyninetybeanpod\\
        \tsix & \numtastsix & \numapitsix & \numgpapitsix & \numlibcapitsix & \numteeapitsix & \maxbbstsix & \reachablegplibcstdtsix{}\% & \numgreedyninetytsix\\
        \midrule
        All & \numtasall & \numapiall & \numgpapiall & \numlibcapiall & \numteeapiall & \maxbbsall & \reachablegplibcstdall{}\% & \numgreedyninetyall\\
        \bottomrule
    \end{tabular}
    \caption{The results of the greedy HLE study. (*) Percentage of reachable basic blocks after implementing 
    all GP and libc APIs. (\dag) The number of TEE-specific APIs needed to achieve 90\% potentially reachable 
    basic blocks (BBs).}
    \label{tab:feas}
\end{table*}

\begin{figure*}[h]
    \definecolor{matplotliborange}{rgb}{1.0, 0.647, 0.0}
    \definecolor{matplotlibgreen}{rgb}{0.0, 0.5, 0.0}
    \definecolor{matplotlibblue}{rgb}{0.0, 0.0, 1.0}
    \begin{tikzpicture}
        \node[rotate=90, anchor=west] at (-3,-0.25) [xshift=-1cm, yshift=-1cm] {\footnotesize{Reachable BBs \%}};
    
    

        \node[anchor=center] (all) at (0.3,0) {
            \includegraphics[width=2.9cm]{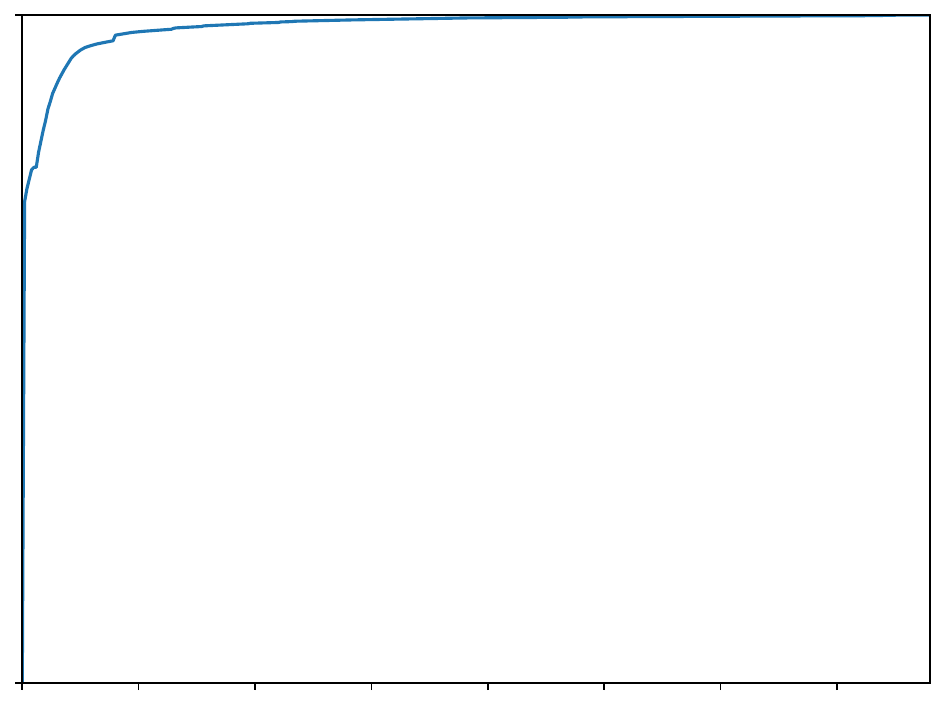}
        };
        \node at (all.north) [yshift=0.05cm,xshift=0cm] {All};
        \node[] at ([xshift=7pt]all.south west) [] {\scriptsize{0}};
        \node[] at ([xshift=-7pt]all.south east) [] {\scriptsize{\numteeapiall}};
        \node[] at ([xshift=1pt,yshift=7pt]all.south west) [] {\scriptsize{\reachablegplibcstdall}};
        \node[] at ([xshift=-1pt,yshift=-7pt]all.north west) [] {\scriptsize{100}};

        \node[anchor=center] (teegris) at (3.6,0) {
            \includegraphics[width=2.9cm]{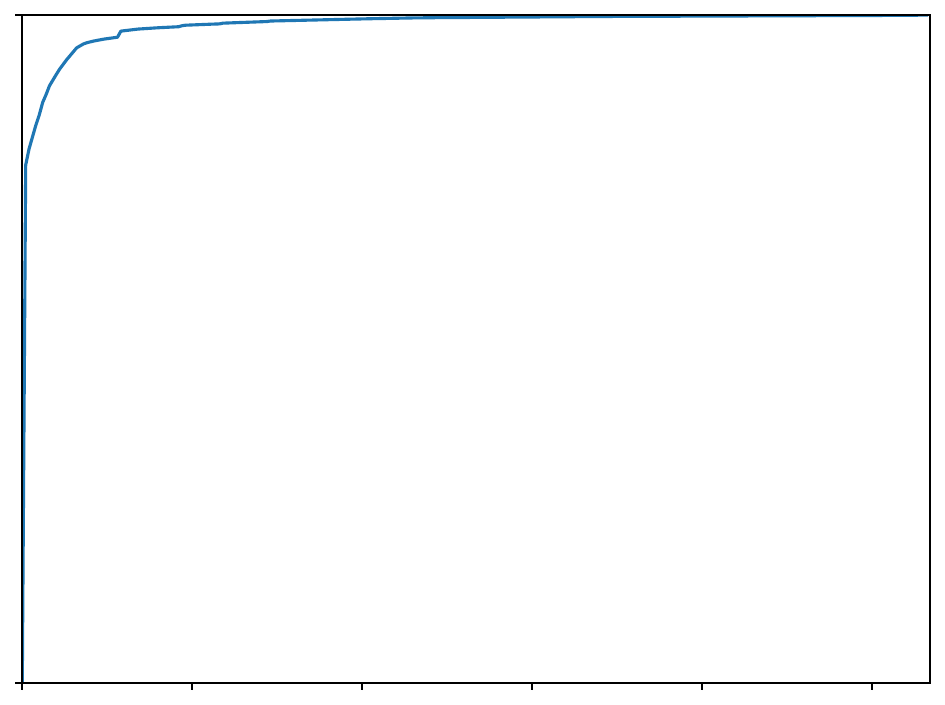}
        };
        \node at (teegris.north) [yshift=0.05cm,xshift=0cm] {\teegris};
        \node[] at ([xshift=7pt]teegris.south west) [] {\scriptsize{0}};
        \node[] at ([xshift=-7pt]teegris.south east) [] {\scriptsize{\numteeapiteegris}};
        \node[] at ([xshift=1pt,yshift=7pt]teegris.south west) [] {\scriptsize{\reachablegplibcstdteegris}};
        \node[] at ([xshift=-1pt,yshift=-7pt]teegris.north west) [] {\scriptsize{100}};

        \node[anchor=center] (mitee) at (6.9,0) {
            \includegraphics[width=2.9cm]{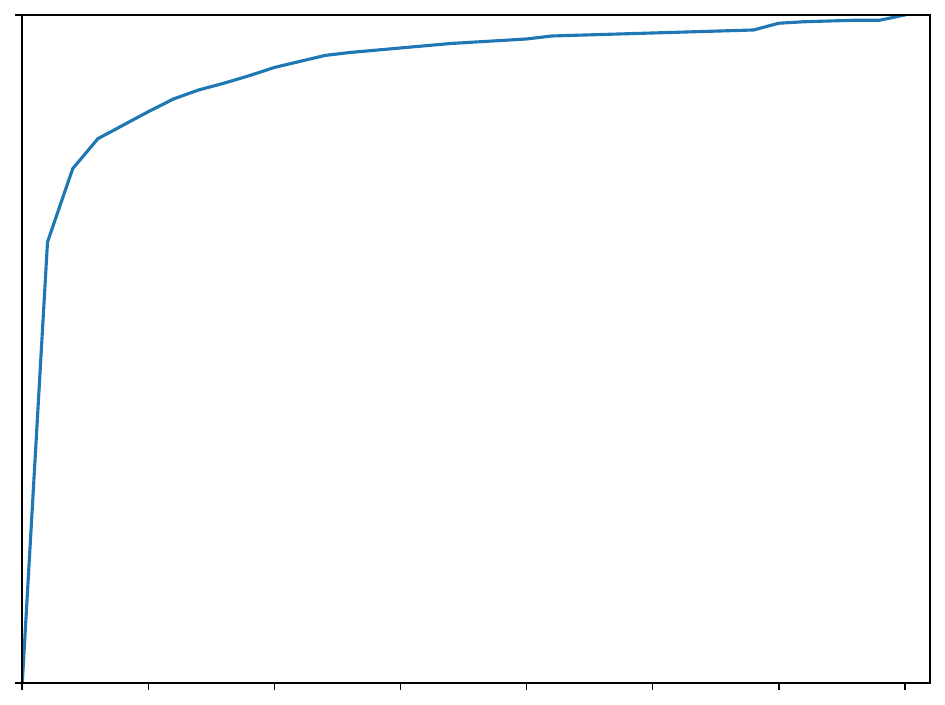}
        };
        \node at (mitee.north) [yshift=0.05cm,xshift=0cm] {\mitee};
        \node[] at ([xshift=7pt]mitee.south west) [] {\scriptsize{0}};
        \node[] at ([xshift=-7pt]mitee.south east) [] {\scriptsize{\numteeapimitee}};
        \node[] at ([xshift=1pt,yshift=7pt]mitee.south west) [] {\scriptsize{\reachablegplibcstdmitee}};
        \node[] at ([xshift=-1pt,yshift=-7pt]mitee.north west) [] {\scriptsize{100}};

        \node[anchor=center] (beanpod) at (10.2,0) {
            \includegraphics[width=2.9cm]{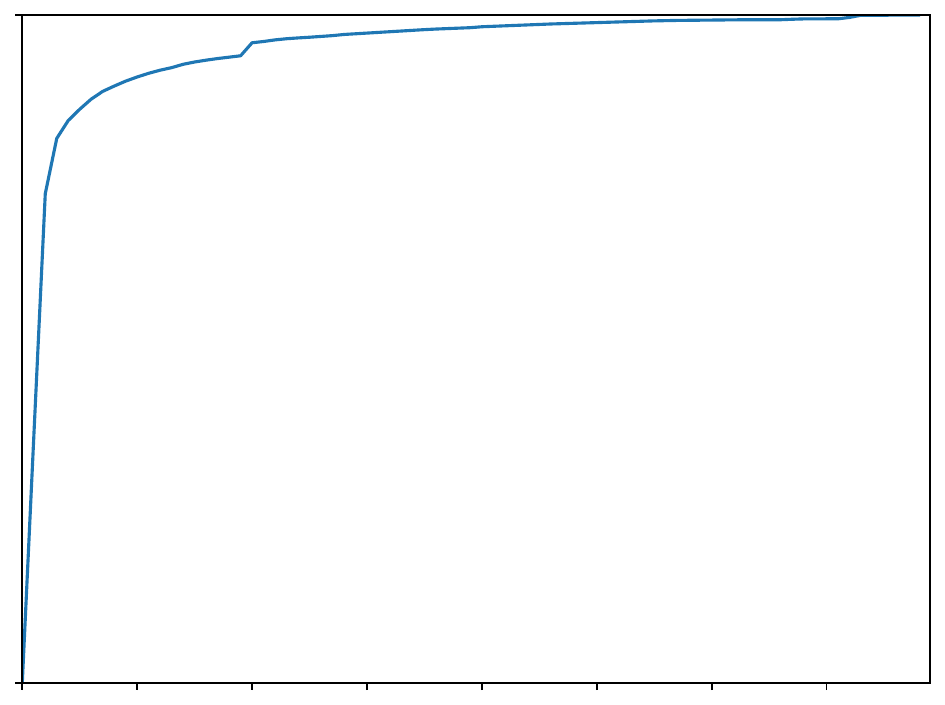}
        };
        \node at (beanpod.north) [yshift=0.05cm,xshift=0cm] {\beanpod};
        \node[] at ([xshift=7pt]beanpod.south west) [] {\scriptsize{0}};
        \node[] at ([xshift=-7pt]beanpod.south east) [] {\scriptsize{\numteeapibeanpod}};
        \node[] at ([xshift=1pt,yshift=7pt]beanpod.south west) [] {\scriptsize{\reachablegplibcstdbeanpod}};
        \node[] at ([xshift=-1pt,yshift=-7pt]beanpod.north west) [] {\scriptsize{100}};

        \node[anchor=center] (tsix) at (13.5,0) {
            \includegraphics[width=2.9cm]{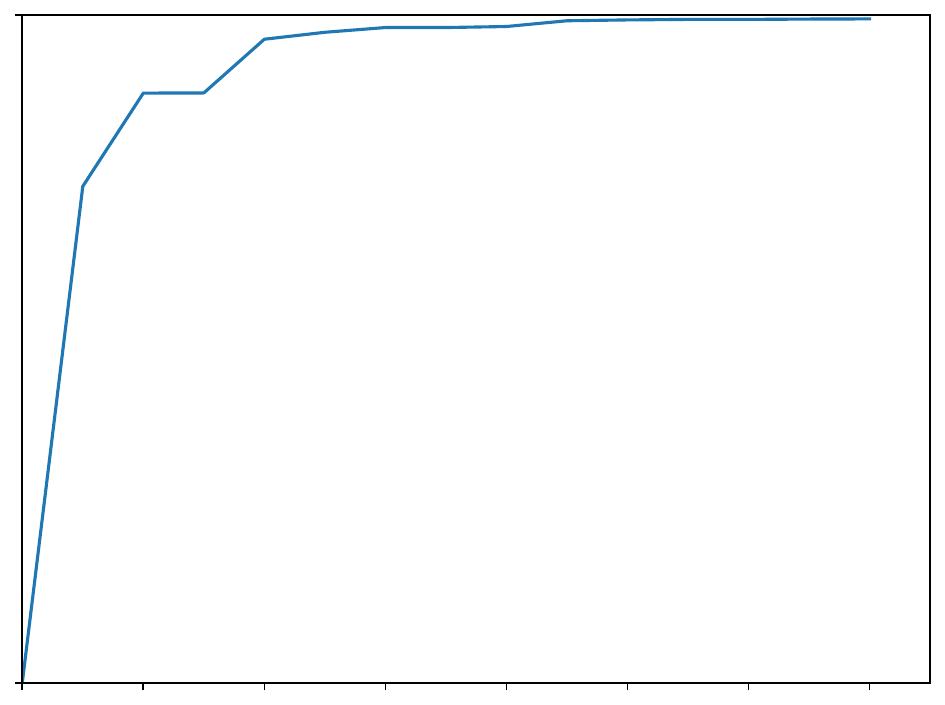}
        };
        \node at (tsix.north) [yshift=0.05cm,xshift=0cm] {\tsix};
        \node[] at ([xshift=7pt]tsix.south west) [] {\scriptsize{0}};
        \node[] at ([xshift=-7pt]tsix.south east) [] {\scriptsize{\numteeapitsix}};
        \node[] at ([xshift=1pt,yshift=7pt]tsix.south west) [] {\scriptsize{\reachablegplibcstdtsix}};
        \node[] at ([xshift=-1pt,yshift=-7pt]tsix.north west) [] {\scriptsize{100}};

        \node[] at (6.75,-1.75) [] {\footnotesize{\# Implemented TEE APIs}};

    \end{tikzpicture}
    \caption{
        The reachable basic blocks in relation to the number of implemented TEE-specific APIs. 
        The graph starts from all GP and libc APIs implemented, and TEE-specific APIs are added 
        with greedy HLE.
    }
    \label{fig:feasibility}
\end{figure*}

\begin{algorithm}[]
\footnotesize
\KwIn{ICFG $\mathcal{G}$(root,B,E); List of TEE APIs $\mathcal{A}$;}
\KwOut{Greedy API list $\mathcal{I}$}

\BlankLine
\textbf{Initialize:} $\mathcal{I} \leftarrow \emptyset$; \\

\While{$size(\mathcal{I})<size(\mathcal{A})$}{
    \ForEach{$a \in \mathcal{A} \setminus \mathcal{I}$}{
        $gain[a] \leftarrow$ $\text{ReachableBlocks}(\mathcal{G},a \cup \mathcal{I}, \mathcal{A})$; 
    }
    $a^* \leftarrow \arg\max_{a} gain[a]$ \;
    $\mathcal{I} \leftarrow \mathcal{I} \cup \{a^*\}$ \;
}
\Return{$\mathcal{I}$}

\BlankLine
\textbf{Procedure} ReachableBlocks($\mathcal{G}$,$\mathcal{I}$,$\mathcal{A}$): \\
\Indp
\ForEach{$b \in \mathcal{G}.B$}{
    \ForEach{$a \in \mathcal{A} \setminus \mathcal{I}$}{
        \If{b calls a}{
            $\mathcal{G}.B \leftarrow \mathcal{G}.B \setminus b$ 
        }
    }
}
\Indp
\Return{size(descendants($\mathcal{G}$,root))}
\Indm

\caption{Greedy High-Level Emulation (HLE)}
\label{alg:greedyhle}
\end{algorithm}

With overall \numteeapiall{} unique TEE APIs across 82 TAs, reverse engineering 
all these proprietary APIs and faithfully implementing them is prohibitively
expensive. 
Instead, we need a way to prioritize TEE-specific APIs. 

We introduce the notion of greedy HLE, a generic technique for incrementally rehosting 
binary components in emulated environments by selectively modeling high-level interfaces. 
The approach leverages two key properties: 1) the greedy-choice property—where modeling the 
next interface with the highest expected benefit yields immediate and independent progress—and 
2) optimal substructure—where each successful emulation step expands the execution context without 
invalidating previous work.
Greedy HLE operates over a well-defined abstraction interface that captures interactions with the 
runtime environment (e.g., TEE-specific API functions). This abstraction provides a uniform representation 
for analyzing, ranking, and selecting the next functions to emulate, independent of their underlying implementation details.
See Algorithm~\ref{alg:greedyhle} for the pseudo code of greedy HLE.

We apply greedy HLE to TEEs by targeting the TEE-specific APIs used by TAs. At each step, we 
select the next API function to implement based on its expected coverage yield—i.e., the number of additional 
basic blocks it unlocks across the TA corpus. By prioritizing APIs that enable the execution of multiple TAs 
or uncover deeper code paths, this strategy rapidly expands the dynamic analysis surface while minimizing 
emulation effort. Although demonstrated in the context of TAs, greedy HLE is a general approach applicable to 
other domains where execution hinges on incrementally modeling high-level interfaces.

To guide this selection process, greedy HLE leverages an interprocedural control flow graph (icfg) to compute 
expected coverage as a function of the currently implemented APIs. The icfg encompasses all relevant code of the 
target, with each basic block annotated by the API calls it makes through the abstraction interface. A basic block 
calling an unimplemented API terminates emulation and is thus removed from the icfg, along with all its incoming 
and outgoing edges. Once all APIs called by a basic block are implemented, the block and its edges are reintroduced.

Starting from an empty list of implemented TEE-specific APIs, greedy HLE incrementally appends to this list by ranking and selecting 
API functions through the abstraction interface according to their coverage contribution—i.e., the number of newly 
reachable basic blocks in the icfg. This structured ranking process ensures systematic and measurable progress toward 
full emulation coverage. 

\textbf{Greedy HLE for TAs.} We conduct a study on the TAs of 
the four fully GP-compliant TEEs, excluding \qsee and \kinibi due to low GP compliance,
to demonstrate how greedy HLE enables emulating TAs from a wide variety of TEEs 
with minimal manual effort. 

For TAs, the relevant code is all the code reachable 
from the GP entrypoint functions since they may be invoked from the normal world.
For each TA, we use Ghidra~\cite{ghidra} to build the icfg starting from 
the GP entrypoint APIs, annotating basic blocks that invoke API functions (GP, libc, or TEE-specific). 
To categorize an API call, we compare its name against the list of APIs defined by the GP Internal Core API and by libc. 
If neither of the lists contain the API name, we classify the call as TEE-specific.

Since our goal is to have a global ranking of the most important TEE-specific APIs, we have to reason across 
our entire dataset of TAs, i.e., we need to apply greedy HLE across multiple TAs. 
We first generate a TEE-level icfg, which includes all relevant TA code within a given TEE, 
by merging the individual TA icfgs belonging to that TEE. Specifically, we create a synthetic "TEE root" basic block that 
connects to the first basic block of each GP entrypoint function from all TA icfgs in that TEE.
This TEE-level icfg allows us to apply greedy HLE across all TAs within a single TEE by measuring the basic blocks reachable 
from the synthetic TEE root. To extend greedy HLE to our entire dataset, we create a global icfg whose synthetic root basic 
block connects to each TEE root basic block.

We first leverage the icfg to measure the amount of code covered by implementing the standard APIs (GP and libc).  
Afterwards, we use greedy HLE to add TEE-specific APIs.
See \autoref{fig:feasibility} 
for the results for each TEE and merged across all TEEs, and 
\autoref{tab:feas} for an overview of the numbers.

The plots show that there are a few high-impact TEE-specific APIs, followed by a long tail of TEE-specific APIs that only marginally 
contribute to the number of potentially reachable basic blocks. 
These results demonstrate that an HLE approach for TAs is feasible.
An emulator supporting the GP APIs and libc APIs achieves around 39\% of potentially reachable 
basic blocks. The reason for this low percentage is \teegris, where the TEE API \texttt{TEES\_IsREESharedMemory} blocks most 
code, resulting in only \reachablegplibcstdteegris\% potentially reachable basic blocks for \teegris. 
This API essentially reimplements an existing GP API (used to check permissions on shared memory). 
After emulating this API, the reachable basic blocks jump to 79\% for \teegris and 82\% for all.
Across all TEEs, implementing ten TEE-specific APIs, guided by greedy HLE, achieves 90\% potentially reachable basic blocks.     

\begin{tcolorbox}
    Leveraging greedy HLE, we identify a limited number of high-impact TEE-specific APIs, which 
    enable emulation of up to 90\% of basic blocks in \sysname.
\end{tcolorbox}

To further motivate \sysname's HLE approach, we study the manual rehosting effort incurred by TEE-specific 
APIs. We manually analyze the top ten TEE-specific APIs for each TEE (in terms of newly reachable basic blocks in the icfg 
after implementation) to try and understand if the TEE-specific API could be replaced by an equivalent 
GP or libc function. We analyze 40 TEE-specific APIs. For 24 of these, there exists an appropriate GP API. 
An example is \texttt{ut\_pf\_cp\_rd\_random} used by \beanpod, which generates random data and can be replaced by 
\texttt{TEE\_GenerateRandom}. For a further five APIs, there exists a corresponding libc API. Overall, 70\% of 
analyzed TEE-specific APIs are replaceable by a GP and libc API. This indicates that in the future, these 
TEE-specific APIs may be replaced by appropriate GP/libc APIs. Furthermore, the rehosting effort for such APIs 
is minimal, as existing functionality can be reused. 
The complete table of API functions can be found in \autoref{tab:customtogp}.

\begin{tcolorbox}
    Most high-impact TEE-specific APIs can be feasibly replaced by existing GP/libc functionality.
\end{tcolorbox}

\begin{table*}[h]
    \scriptsize
    \centering
    \begin{tabular}{lllll}
        TEE&
        API & 
        Functionality &
        GP equivalent &
        libc equivalent \\
        \midrule
        \teegris & \texttt{TEES\_IsREESharedMemory} & check memory permissions & \texttt{TEE\_CheckMemoryAccessRights} & N/A\\
        \teegris & \texttt{TEES\_GetClientCredentials} & retrieve NWCA authentication info & \texttt{TEE\_GetProperty} (client) & N/A\\
        \teegris & \texttt{sqrf, roundf...} & float math operations & N/A & N/A\\
        \teegris & \texttt{TA\_Communication\_mpos\_check\_iccc} & IPC communication & \texttt{TEE\_InvokeTACommand} & N/A \\
        \teegris & \texttt{hdm\_ICCC\_check} & IPC communication & \texttt{TEE\_InvokeTACommand} & N/A \\
        \teegris & \texttt{TEES\_UnwrapSecureObject} & Access encrypted persistent data & \texttt{TEE\_ReadPersistentObject} & N/A \\
        \teegris & \texttt{OPENSSL\_malloc} & memory allocation & \texttt{TEE\_Malloc} & N/A\\
        \teegris & \texttt{TEES\_CheckSecureObjectCreator} & check integrity of persistent data & \texttt{TEE\_OpenPersistentObject} & N/A\\
        \teegris & \texttt{TEES\_SPIWriteRead} & peripheral access & N/A & N/A\\
        \teegris & \texttt{TEES\_DeriveKeyKDF} & generate key & \texttt{TEE\_GenerateKey} & N/A\\
        \mitee & \texttt{tee\_se\_open\_spi\_clk} & peripheral access & N/A & N/A\\
        \mitee & \texttt{tee\_get\_enc\_rot}  & retrieve TEE specific value & \texttt{TEE\_GetProperty} & N/A\\
        \mitee & \texttt{tee\_get\_cpuid} & retrieve device specific value & \texttt{TEE\_GetProperty} & N/A\\
        \mitee & \texttt{TEE\_SEChannelClose} & close IPC communication & \texttt{TEE\_CloseTASession} & N/A\\
        \mitee & \texttt{TEE\_SaveTA\_Data} & write data to persistent storage & \texttt{TEE\_WriteObjectData} & N/A\\
        \mitee & \texttt{gen\_random} & generate random data & \texttt{TEE\_GenerateRandom} & N/A\\
        \mitee & \texttt{tee\_se\_close\_spi\_clk} & peripheral access & N/A & N/A\\
        \mitee & \texttt{TEE\_SESessionOpenBasicChannel} & IPC communication & \texttt{TEE\_OpenTASession} & N/A\\
        \mitee & \texttt{TEE\_SEChannelGetNumber\_ext} & IPC communication & N/A & N/A\\
        \mitee & \texttt{TEE\_SESessionOpenLogicalChannel} & IPC communication & \texttt{TEE\_OpenTASession} & N/A\\
        \beanpod & \texttt{TEE\_LogPrintf} & logging & N/A & \texttt{printf}\\
        \beanpod & \texttt{msee\_ta\_printf\_va} & logging & N/A & \texttt{printf}\\
        \beanpod & \texttt{ut\_pf\_log\_msg} & logging & N/A & \texttt{printf}\\
        \beanpod & \texttt{ut\_pf\_km\_get\_hmac\_key} & retrieve keymaster key & \texttt{TEE\_InvokeTACommand} & N/A\\
        \beanpod & \texttt{mdrv\_open} & initialize session with driver & \texttt{TEE\_OpenTASession} & N/A\\
        \beanpod & \texttt{ut\_pf\_cp\_rd\_random} & generate random data & \texttt{TEE\_GenerateRandom} & N/A\\
        \beanpod & \texttt{base64\_decode} & base64 decoding & N/A & N/A\\
        \beanpod & \texttt{TEE\_RpmbOpenSession} & open object from RPMB & \texttt{TEE\_OpenPersistentObject} & N/A\\
        \beanpod & \texttt{TEE\_RpmbReadData} & read data from RPMB & \texttt{TEE\_ReadObjectData} & N/A\\
        \beanpod & \texttt{ut\_pf\_info\_get\_deviceinfo} & get device information & \texttt{TEE\_GetProperty} & N/A\\
        \tsix & \texttt{debug\_log} & logging & N/A & \texttt{printf}\\
        \tsix & \texttt{debug\_log2} & logging & N/A & \texttt{printf}\\
        \tsix & \texttt{\_\_assert\_fail} & log and abort & \texttt{TEE\_Panic} & N/A\\
        \tsix & \texttt{TEE\_GetBootSeed} & retrieve hw specific data & \texttt{TEE\_GetProperty} & N/A\\
        \tsix & \texttt{platform\_spi\_write\_read} & peripheral access & N/A & N/A\\
        \tsix & \texttt{init\_ta\_session} & setup IPC & \texttt{TEE\_OpenTASession} & N/A\\
        \tsix & \texttt{sf\_spi\_write\_buf} & peripheral access & N/A & N/A\\
        \tsix & \texttt{sf\_spi\_write\_then\_read\_buf} & peripheral access & N/A & N/A\\
        \tsix & \texttt{check\_license} & get device information & \texttt{TEE\_GetProperty} & N/A\\
        \tsix & \texttt{platform\_open\_driver} & intitiate IPC & \texttt{TEE\_OpenTASession} & N/A\\
        \bottomrule
    \end{tabular}
    \caption{The analyzed highest-impact TEE-specific APIs, their reverse-engineered functionality, and the possible replacement candidates from libc or GP.}
    \label{tab:customtogp}
\end{table*}

\section{Implementation}\label{sec:implementation}

We implement \sysname on top of Qiling~\cite{qiling}, which itself is a wrapper 
around Unicorn~\cite{unicorn}. in around \numlines lines of Python. 

\textbf{Annotation of Addresses.}
\sysname requires the addresses of API functions to enable API-level interception.

\teegris and \beanpod TAs are dynamically linked ELF files that export/import the GP, libc, 
and TEE APIs. Hence, the loader can simply parse the ELF header to determine the
addresses of these functions. \tsix and \mitee TAs are
statically compiled and, thus, require special handling.  For these TAs, the
loader uses a TA-specific configuration file, which stores the offsets to the
relevant functions. We automate the generation of this file using a Ghidra
headless script.  The script uses logging strings to map functions to the GP
and libc APIs.  The script further annotates call sites that have not
been marked as an API but contain unhandled system calls.  These unhandled
system calls require a one-time effort to manually complete the previously
generated configuration file to annotate the API making the system call in
question.  While the automation with Ghidra could be further improved using
binary similarity or similar techniques, we consider the manual effort
negligible.

\textbf{API Hooking.}
To hook the API calls for dynamically linked TAs, \sysname overwrites the global 
offset table entries with unique pointers. When the program counter in the 
emulator core is equal to one of these unique pointers, the corresponding API handler is called.
For statically linked TAs, \sysname parses the configuration file and hooks the 
inline API function addresses.

The API handlers are implemented in Python and use the Qiling API to update 
registers and memory. Listing~\ref{lst:api} shows the implementation for the \texttt{TEE\_MemMove} 
function. The implementation first retrieves the function arguments. It uses \sysname's 
address sanitizer (ASAN) implementation to check if memory accesses are valid. Before copying 
data to the destination, it synchronizes the shared memory. Finally, it copies the 
data from the source and writes it to the destination. The Qiling API is used to 
set the return value. 
To return to the callee position, the address in the \texttt{lr} register is 
restored to the \texttt{pc}.
This simple API hook is able to detect non-crashing out-of-bounds accesses with ASAN, 
simulates shared memory accesses, and correctly implements the \texttt{TEE\_MemMove} 
functionality.

\begin{figure}[]
    \input{listings/api.tex}
    \captionof{lstlisting}{
        The \texttt{TEE\_MemMove} API implementation of \sysname.
    }
    \label{lst:api}
\end{figure}

\textbf{Fuzzing.}
The fuzzing mode uses AFL++~\cite{AFLplusplus}'s Unicorn mode in combination with 
Qiling's \texttt{libunicornafl} support to fuzz the target TA. 
In fuzzing mode, the emulator acts as the target program for AFL++.
To start the fuzzing mode, the user specifies the target TA and a Python harness 
script as arguments to the emulator. The TA manager 
loads the harness script. The harness specifies two functions.
The \texttt{init} function, which runs before the fuzzing loop and can be used 
to set up a specific state in the target TA. The \texttt{init} function is called 
before AFL's forkserver is started, so it only runs once. The second harness function, \texttt{place\_input\_callback}, 
is invoked for every fuzzing iteration and is responsible 
for deserializing the fuzzer-provided bytes into the arguments of the 
\texttt{TA\_InvokeCommandEntryPoint} function.

\textbf{Address Sanitization (ASAN).}
Since \sysname hooks the memory management APIs (such as \texttt{TEE\_Malloc} and
\texttt{TEE\_Free}), \sysname leverages these hooks to track the heap state. When allocating 
memory, \sysname sets up a redzone around the returned memory chunk and hooks 
accesses to this region to detect out-of-bounds accesses. It also tracks the state of chunks 
to detect invalid frees and double frees. Other APIs that read or write from or to memory 
use the ASAN sanitizer to ensure memory accesses are valid.

\textbf{Emulating Common TEE Vulnerabilities.}

\sysname ensures TA-specific vulnerabilities are correctly emulated.
Specifically, parameter type-confusion bugs~\cite{globalconfusion} and
time-of-check to time-of-use (TOCTTOU) bugs can be faithfully emulated with \sysname.

Parameter type-confusion bugs are the result of improper sanitization of untrusted input. 
These inputs can lead to vulnerabilities allowing an attacker to arbitrarily control pointers 
in the virtual address space of the affected TA.
The TA manager ensures that these type confusions are triggerable by handling the client's inputs 
faithfully using \optee as the reference implementation. In~\autoref{sec:debug}, we show 
how we use the emulator to debug and exploit a n-day parameter type-confusion bug.

TOCTTOU bugs in the context of TAs arise from memory being shared between the normal and secure world. 
If the TA operates directly on the shared memory, concurrent changes by the normal world can lead to 
inconsistent or stale reads — i.e., the TA checks a value and later uses it after the normal world has modified it, 
opening a race window that an attacker can exploit. \sysname models this shared-memory interaction precisely: 
changes by the NWCA to the TA memory buffer are propagated to the emulator via Linux shared memory, 
which allows concurrent modifications by the NWCA. In~\autoref{sec:fidelity} we 
demonstrate how we are able to reproduce an n-day TOCTTOU bug with \sysname.

\section{Evaluation}\label{sec:evaluation}

We evaluate \sysname's emulation capabilities across various TEEs, first assessing its
fidelity and then focusing on the use cases: fuzzing and debugging.

Throughout the evaluation, we aim to answer the following research questions:
\begin{description}

        \item[\textbf{RQ1}] Can \sysname effectively be used to reproduce known (n-day) vulnerabilities in commercial TAs?
        
        \item[\textbf{RQ2}] Can \sysname effectively discover new (0-day) vulnerabilities in commercial TAs, reproducible on-device?

        \item[\textbf{RQ3}] Can \sysname unlock fuzzing capabilities for commercial TAs?

\end{description}

\subsection{Fidelity (RQ1 \& RQ2)}\label{sec:fidelity}
To evaluate the fidelity of \sysname, we first evaluate if \sysname can be used
to reproduce n-day vulnerabilities. In the second part, we evaluate whether the
crashes in TAs found during development and fuzzing can be reproduced on our
physical commercial off-the-shelf (COTS) testing devices.

\begin{table*}[t]
    \small
    \centering
    \begin{tabular}{llllc}
        TEE&
        TA &
        CVE-ID &
        Vuln Type  &
        Reproduced \\
        \midrule
        \teegris & 00000000-0000-0000-0000-000000000046 & SVE-2019-14867 & parameter type confusion & \checkmark \\
        \teegris & 00000000-0000-0000-0000-000048444350 & SVE-2019-14850 & parameter type confusion & \checkmark \\
        \teegris & 00000000-0000-0000-0000-000048444350 & CVE-2019-20545 & stack buffer overflow & \checkmark \\
        \beanpod & d78d338b1ac349e09f65f4efe179739d.ta & CVE-2020–14125 & heap buffer overflow & \checkmark \\
        \beanpod & 08110000000000000000000000000000.ta & CVE-2023-32835 & parameter type confusion & \checkmark \\
        TrustedCore & task\_storage & N/A & TOCTTOU & \checkmark \\
        \bottomrule
    \end{tabular}
    \caption{The results of reproducing n-day vulnerabilities with \sysname.}
    \label{tab:oldrepro}
\end{table*}

For the first fidelity study, we pick \numndays n-day TA vulnerabilities from \sysname's
supported TEEs. Additionally, we added support to \sysname for emulating a
single TrustedCore Huawei TA to demonstrate the reproducibility of shared
memory TOCTTOU vulnerabilities. We chose vulnerabilities for which a proof of
concept or a writeup exists, allowing us to write our own proof of concept
NWCA. The vulnerabilities we chose include common memory corruptions, i.e.,
overflows and TEE-specific vulnerabilites such as parameter type-confusions and
TOCTTOUs. We compile and run our proof of concept NWCA against the vulnerable
TA emulated in \sysname. We mark this as a success if \sysname detects the
crash.  \autoref{tab:oldrepro} shows the n-day vulnerabilities and the result
of replaying the proof of concept against the emulator. We successfully
reproduced all vulnerabilities in \sysname. 

The second fidelity study aims to reproduce 0-day vulnerabilities found during
development of or fuzzing with \sysname on physical COTS devices. For each of
these vulnerabilities, we write a proof of concept NWCA, compile them for the
corresponding Android device, and check if we can trigger the crash on the
device. For \beanpod TAs, we use a Xiaomi Redmi Note 11s. For \mitee TAs we use
a Xiaomi Redmi Note 13 5G and for the \tsix TAs we use a Ulephone Power Armor
18.  All our testing devices are rooted to allow interaction with the TEE
kernel driver.  Note that on-device, we do not have any introspection
capabilities and can only detect if the TA crashed by observing the return code
of \texttt{TEEC\_InvokeCommand} (we mark a crash as reproduced if we observe a
return value of \texttt{TEE\_ERROR\_TARGET\_DEAD} and a \texttt{returnOrigin}
of \texttt{TEEC\_ORIGIN\_TEE}). \autoref{tab:vulns} lists the vulnerabilities
discovered with \sysname. The column "Reproduced on Device" marks whether we
were able to reproduce the vulnerability on the device.  Out of the \numvulns
vulnerabilities we reproduced \numvulnsdevicerepro. For
\numvulnsnotondevicerepro vulnerabilities, we could not reproduce them due to
our testing devices not supporting that TA. Due to the TEE's enforcement of
only running correctly signed code, it is not possible to load TAs from another
device onto our testing devices.  We were unable to reproduce a 4-byte heap
overflow vulnerability due to the limited corruption primitive, and the
underlying allocator mechanism.

\subsection{Fuzzing TAs (RQ3)}

\begin{table*}[t]
    \small
    \centering
    \begin{tabular}{llllc}
        TEE&
        TA &
        \makecell{Found During} &
        Vuln Type  &
        \makecell{Reproduced\\on Device} \\
        \midrule
        \beanpod & 08110000000000000000000000000000.ta & development & 4 byte heap overflow & \ding{55} \\       
        \beanpod & df1edda8627911e980ae507b9d9a7e7d.ta & development & parameter type confusion & \checkmark\\
        \mitee & 377ee4e8-af0e-474f-a9d636a9268fe85c.ta & development & TOCTTOU & \checkmark \\
        \mitee & 377ee4e8-af0e-474f-a9d636a9268fe85c.ta & fuzzing & stack overflow & \checkmark \\
        \mitee & 377ee4e8-af0e-474f-a9d636a9268fe85c.ta & fuzzing & heap overflow & \checkmark \\
        \mitee & f13010e0-2ae1-11e5-896a0002a5d5c51d.ta & fuzzing & buffer underflow & \checkmark \\
        \mitee & a734eed9-d6a1-4244-aa507c99719e7b7f.ta & development & parameter type confusion & \ding{55}* \\
        \mitee & a734eed9-d6a1-4244-aa507c99719e7b7f.ta & fuzzing & null pointer dereference & \ding{55}* \\
        \mitee & a734eed9-d6a1-4244-aa507c99719e7b7f.ta & fuzzing & oob. read & \ding{55}* \\
        \mitee & 9811c1f6-47e3-5cea-ae6ef62ba433c4fd.ta & fuzzing & oob. read & \ding{55}* \\
        \mitee & 3d08821c-33a6-11e6-a1fa089e01c83aa2.ta & fuzzing & double free & \checkmark \\
        \mitee & 59a4867c-9fe5-f7c2-b409a46bae6ff73e.ta & fuzzing & oob. read & \ding{55}* \\
        \mitee & 8aaaf201-2460-0000-7143fe4f7c823c80.ta & fuzzing & arb. read & \checkmark  \\
        \mitee & 8aaaf201-2460-0000-7143fe4f7c823c80.ta & fuzzing & stack overflow & \checkmark  \\
        \mitee & 86f623f6-a299-4dfd-b560ffd3e5a62c29.ta & fuzzing & arb. read & \checkmark  \\
        \tsix & 9459b61a-02d3-4d1e-b68be94397e7ca8c.ta & development & parameter type confusion & \checkmark  \\
        \tsix & 9459b61a-02d3-4d1e-b68be94397e7ca8c.ta & fuzzing & data section overflow & \checkmark  \\
        \bottomrule
    \end{tabular}
    \caption{
    The vulnerabilities found with \sysname either during development or fuzzing. 
    (*)These vulnerabilities could not be reproduced due to our testing devices not supporting the TA.
    }
    \label{tab:vulns}
\end{table*}

\begin{figure}[]
    \input{listings/harness.tex}
    \captionof{lstlisting}{
        An example harness for the \texttt{14b0aad8-c011-4a3f-b66aca8d0e66f273.ta} TA.
        The \texttt{place\_input\_callback} function is called for every fuzzing iteration 
        with the fuzzer-generated bytes in the \texttt{input} argument. The TA only 
        exposes 5 commands and expects a \texttt{0x608}-sized output buffer as the 
        second parameter.
    }
    \label{lst:harness}
\end{figure}

\begin{figure*}[t]
    \definecolor{matplotliborange}{rgb}{1.0, 0.647, 0.0}
    \definecolor{matplotlibgreen}{rgb}{0.0, 0.5, 0.0}
    \definecolor{matplotlibblue}{rgb}{0.0, 0.0, 1.0}
    \begin{tikzpicture}
        \node[rotate=90, anchor=west] at (-3,-0.5) [xshift=-1cm, yshift=-1cm] {\footnotesize{Basic Block Coverage \%}};
    
        \node[] at (-1.5,1.1) [] {\footnotesize{100}};
        \node[] at (-1.45,0) [] {\footnotesize{50}};
        \node[] at (-1.4,-1.1) [] {\footnotesize{0}};
    
        \node[anchor=center] (all) at (0.3,0) {
            \includegraphics[width=3.1cm]{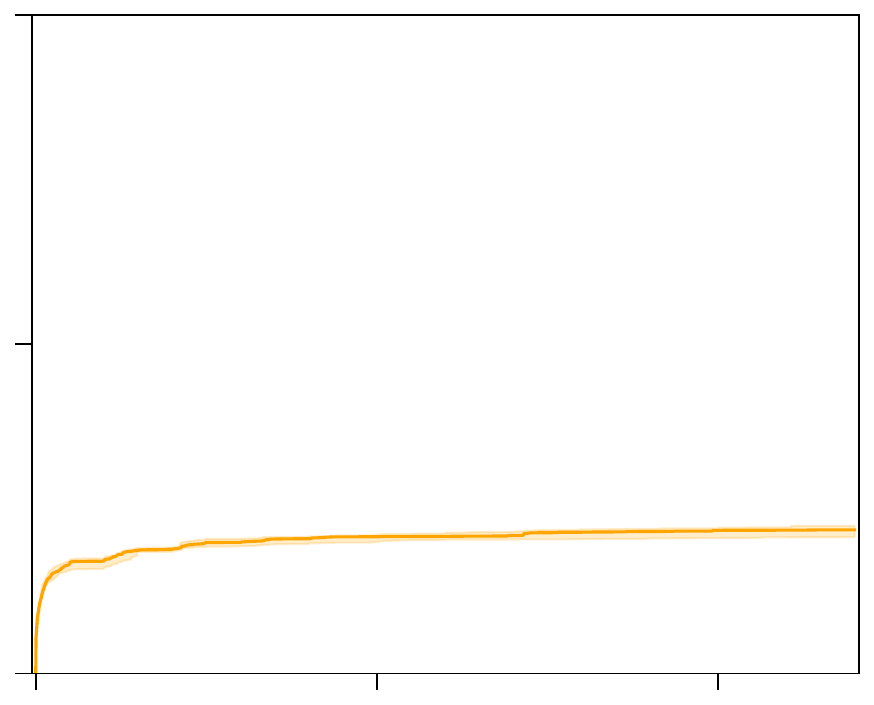}
        };
        \node at (all.north) [yshift=0.05cm,xshift=0cm] {All};

        \node[anchor=center] (teegris) at (3.6,0) {
            \includegraphics[width=3.1cm]{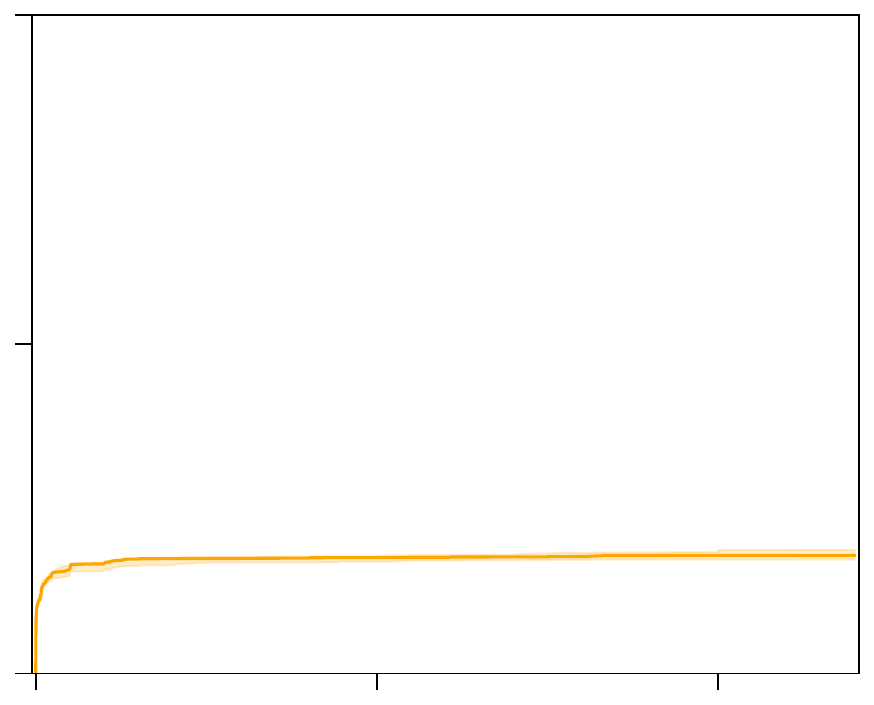}
        };
        \node at (teegris.north) [yshift=0.05cm,xshift=0cm] {\teegris};

        \node[anchor=center] (mitee) at (6.9,0) {
            \includegraphics[width=3.1cm]{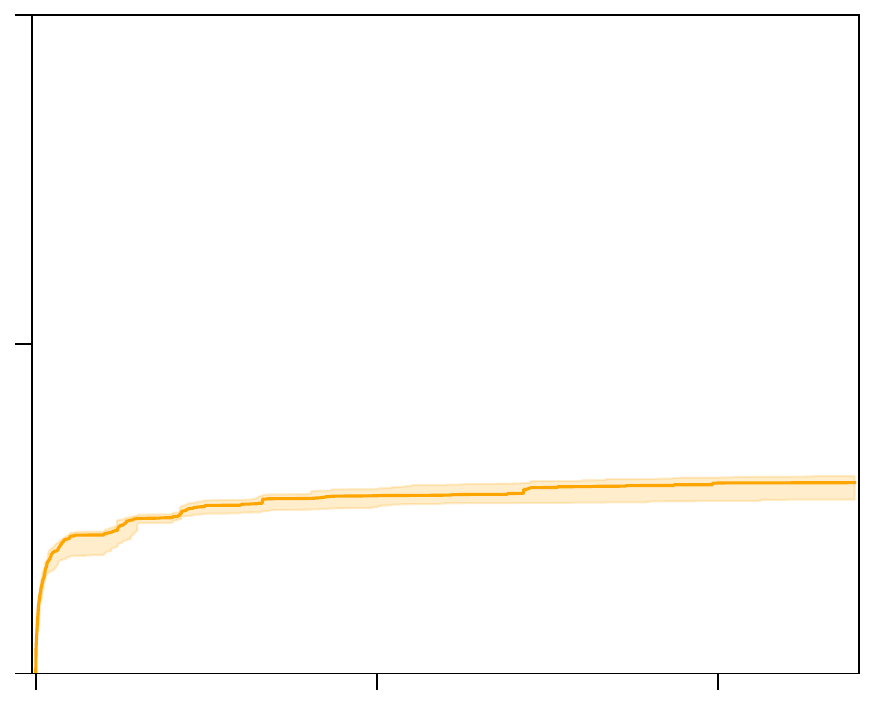}
        };
        \node at (mitee.north) [yshift=0.05cm,xshift=0cm] {\mitee};

        \node[anchor=center] (beanpod) at (10.2,0) {
            \includegraphics[width=3.1cm]{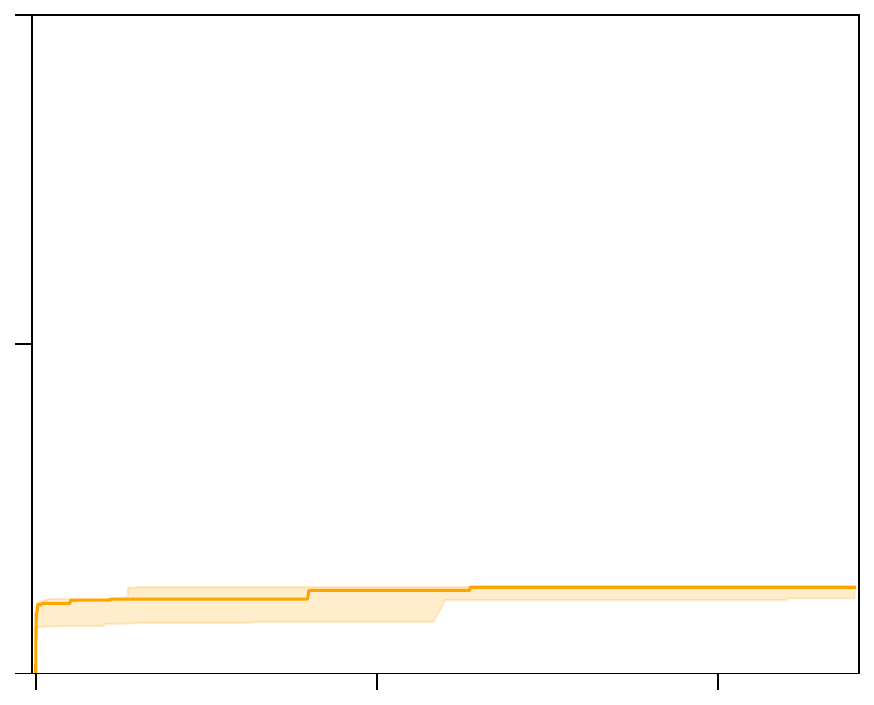}
        };
        \node at (beanpod.north) [yshift=0.05cm,xshift=0cm] {\beanpod};

        \node[anchor=center] (tsix) at (13.5,0) {
            \includegraphics[width=3.1cm]{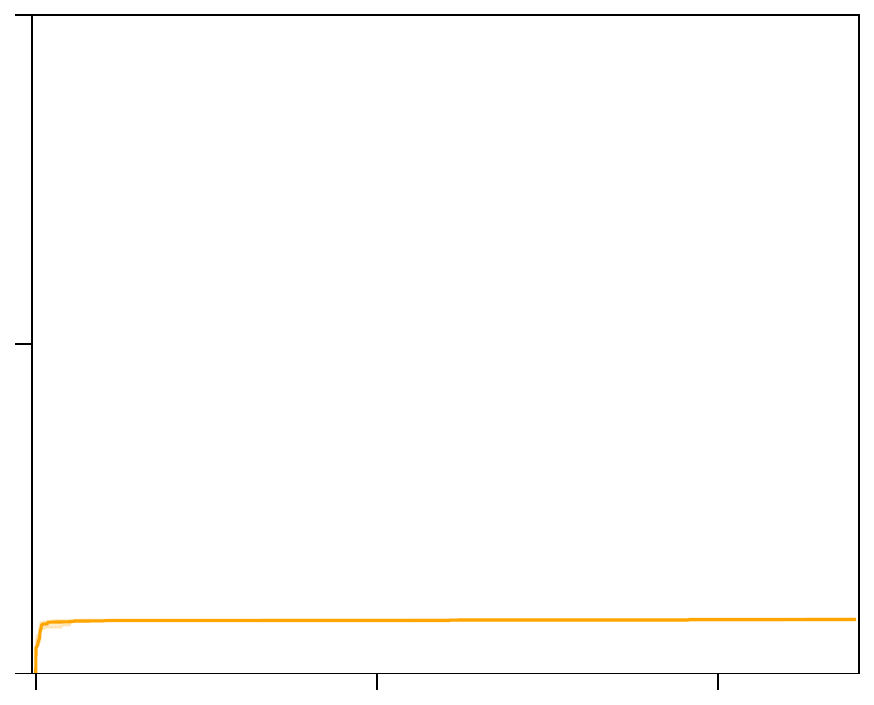}
        };
        \node at (tsix.north) [yshift=0.05cm,xshift=0cm] {\tsix};

        \node[] at (6.75,-1.5) [] {\footnotesize{Time: 0 to 24 hours}};

    \end{tikzpicture}
    \caption{
        Basic block coverage observed after fuzzing \numfuzztas TAs five times for 24 hours as a percentage of the 
        reachable basic blocks. 
    }
    \label{fig:fuzz}
\end{figure*}

\begin{table}[]
    \scriptsize
    \centering
    \begin{tabular}{lrrrrrr}
        TEE&
        \makecell{\# Fuzzed\\TAs} &
        \# Crash &
        \makecell{\# Bug\\Crash} &
        \makecell{\# Not impl.\\API Crash} &
        \makecell{\# max.\\BBs} &
        \makecell{\# cov.\\BBs} \\
        \midrule
        \teegris & \numfuzztasteegris & \numfuzzcrashesteegris & \numfuzzbugteegris & \numfuzznotimplteegris & \numfuzzmaxbbteegris & \numfuzzbbteegris \\
        \mitee & \numfuzztasmitee & \numfuzzcrashesmitee & \numfuzzbugmitee & \numfuzznotimplmitee & \numfuzzmaxbbmitee & \numfuzzbbmitee \\
        \beanpod & \numfuzztasbeanpod & \numfuzzcrashesbeanpod & \numfuzzbugbeanpod & \numfuzznotimplbeanpod & \numfuzzmaxbbbeanpod & \numfuzzbbbeanpod \\
        \tsix & \numfuzztastsix & \numfuzzcrashestsix & \numfuzzbugtsix & \numfuzznotimpltsix & \numfuzzmaxbbtsix & \numfuzzbbtsix \\
        \midrule
        All & \numfuzztas & \numfuzzcrashes & \numfuzzbug & \numfuzznotimpl & \numfuzzmaxbb & \numfuzzbb \\
        \bottomrule
    \end{tabular}
    \caption{The number of fuzzed TAs, uncovered crashes, and covered basic blocks of our 
            fuzzing campaign.}
    \label{tab:fuzz}
\end{table}

We ran a fuzzing campaign against \numfuzztas{} TAs. We chose these TAs based on the 
expected input format in their \texttt{TA\_InvokeCommand} entrypoint function. 
All of the fuzzed TAs expect a single input buffer. This allows us to apply the same logic 
to all harnesses, effectively targeting the TAs. The other TAs implement a more complicated input format. 
Fuzzing these TAs without a bespoke harness will lead to few meaningful findings, 
as all inputs will be rejected early in the entrypoint function.

The fuzzing logic of the harness 
uses the first byte of the fuzzing data to choose the command and then writes 
the rest of the fuzzing bytes to the input buffer. For each of the \numfuzztas 
TAs, the harness ensures that the parameter types and buffer sizes corresponding to a 
given command are as expected by the TA, such that the 
fuzzer can reach code where the TA starts processing the fuzzing data. 
Listing~\ref{lst:harness} shows an example of such a harness.

We fuzz each TA five times for 24 hours. 
We compare the achieved coverage to all potentially reachable basic blocks from the entrypoints. 
\autoref{fig:fuzz} shows the coverage for the various TEEs. 
In our fuzzing campaign, we modify the default API function, called for non-implemented APIs, 
to cause a crash registered with AFL. We replay all crashes, deduplicate, and then 
triage whether the crashes are true positives or false positives caused by missing 
API handling. \autoref{tab:fuzz} shows the results. Out of the overall \numfuzzcrashes{} crashes, 
\numfuzzbug{} (72\%) are true positive crashes. The remaining \numfuzznotimpl crashes are due to 
the TA calling a non-implemented API.
On average, we achieve 100 executions per second, which is a 5-10 times speedup compared to 
Partemu's reported fuzzing speed~\cite{partemu}.

This fuzzing campaign uncovered \numfuzzvuln vulnerabilities, see \autoref{tab:vulns}. 
As pointed out by related work~\cite{teezz,syncemu}, fuzzing TAs is challenging due to 
state, complex input format, and large input sizes. In our fuzzing campaign, we 
address none of these challenges as we fuzz a single invocation of \texttt{TA\_InvokeCommandEntryPoint} 
in each fuzzing iteration with raw bytes mutated by AFL. Instead, \sysname enables 
setting up a fuzzing campaign with coverage for a large number of TAs from different TEEs.

\subsection{Case Study: Debugging}\label{sec:debug}

Another important use case for \sysname is debugging. We demonstrate how we used 
\sysname to improve an exploit for the parameter type confusion bug (CVE-2023-32835) in the 
08110000000000000000000000000000 \beanpod TA. Exploitation of this vulnerability has 
already been presented in~\cite{globalconfusion}. As the details of the vulnerability 
are known, we instead focus on porting the existing proof of concept to \sysname and adjusting the exploit 
from a stack-based PC hijack to a more reliable GOT overwrite-based PC hijack to build an arbitrary 
read primitive. 

CVE-2023-32835 is a parameter type confusion bug, which causes the TA to write an integer 
at an attacker-controlled address. Listing~\ref{lst:keyinst} shows the relevant 
code. Due to the \texttt{paramTypes} argument not getting checked, an attacker can 
supply arbitrary values, which will be interpreted as pointers by the TA. By supplying 
arbitrary pointers as the \texttt{out} buffer, controlled data from the \texttt{in} 
buffer can be written to chosen addresses. The exploit from~GlobalConfusion~\cite{globalconfusion} triggers 
the arbitrary write by sending a request to the TA with \texttt{keycount} equal to one, 
\texttt{in\_buf[22]} containing the chosen arbitrary write value, and \texttt{out} being the 
chosen arbitrary write location. It then repeatedly triggers this arbitrary write to overwrite 
the return address stored on the stack to jump to shellcode. We demonstrate how using \sysname 
we can exploit this arbitrary write and transform it to an arbitrary memory read. Note that 
directly using the parameter type confusion for an arbitrary read is not feasible due to the magic 
number check in the \texttt{in} buffer.

\begin{figure}[]
    \input{listings/0811.tex}
    \captionof{lstlisting}{
        The simplified code of the parameter type confusion in the 08110000000000000000000000000000
        TA. Since the parameters are not checked, an attacker can write data from the \texttt{in} 
        buffer to an arbitrary pointer supplied with a value in \texttt{out}.
    }
    \label{lst:keyinst}
\end{figure}

We download the existing proof of concept from the artifact of~GlobalConfusion~\cite{globalconfusion}\footnote{\url{https://github.com/HexHive/GlobalConfusion/blob/public/pocs/ta_keyinstall/jni/poc.c}} and recompile it for the emulator. Due to \sysname's 
NWCA compilation pipeline, we only need minimal changes to the existing proof of concept to run 
it against the emulator and reproduce the arbitrary write. The TA uses the TEE-specific \texttt{msee\_ta\_printf\_va} API 
before executing the vulnerable code. Thanks to greedy HLE, this API is supported in \sysname, and all the code 
relevant to the proof of concept can be executed. 
The existing proof of concept overwrites a hard coded stack address to hijack the PC. This way of achieving code 
execution relies on knowledge of the stack address where the return 
address is stored. Instead, we will use \sysname's gdb integration to build an exploit that   
overwrites the GOT to achieve a reliable memory leak. Note that the exploit works because 
all the involved binaries are not position independent, and the TA's GOT is writable.

\begin{figure}[]
    \input{listings/gdb.tex}
    \captionof{lstlisting}{
       The output of gdb, attached to \sysname's gdb server at the first invocation of \texttt{TEE\_CheckMemoryAccessRights}. 
    }
    \label{lst:gdb}
\end{figure}

We run the proof of concept with gdb attached and set a breakpoint on the first call to \texttt{TEE\_CheckMemoryAccessRights}. 
Listing~\ref{lst:gdb} shows the relevant output of the gdb context at the breakpoint.
Register \texttt{r1} holds the pointer to the \texttt{in} 
buffer. More importantly, \texttt{r3} holds \texttt{out\_size}. The proof of concept uses an \texttt{out\_size} of 4. 
Since the TA does not check the parameter types, we can fully control both \texttt{r1} 
and \texttt{r3}. Furthermore, by overwriting the GOT entry of 
\texttt{TEE\_CheckMemoryAccessRights}, we can jump to a chosen address. 

We analyze the TA and the TA's loader \texttt{libld-l4.so} for useful gadgets.
In the loader we find the following gadget: \texttt{ldr r3, [r3] ; str r3, [r1] ; bx lr}, reading 
four bytes from \texttt{r3} and writing them to \texttt{r1}. Since we control both \texttt{r3} 
and \texttt{r1}, we can use this gadget for arbitrary read. Our exploit first overwrites the GOT 
entry of \texttt{TEE\_CheckMemoryAccessRights} with the address of this gadget in the loader. 
Since \sysname itself overwrites GOT entries to 
point to hooked memory, this GOT overwrite works transparently in \sysname.
Afterwards, when invoking the TA with \texttt{out\_size} set to the target arbitrary read 
address and a legitimate \texttt{in} buffer, we can leak memory 4 bytes at a time.

With \sysname's gdb integration and NWCA compilation pipeline, we were able to easily 
port an existing exploit to \sysname, debug it, and adjust the exploit to build a 
100\% reliable memory leak primitive.

\section{Discussion}\label{sec:discussions}

\textbf{Extending \sysname to Non-Compliant TEEs.}
Extending \sysname to non-compliant TEEs requires first understanding 
how the TAs are loaded and invoked by the TEE. 
Further, it requires implementing the TEE-specific APIs.

The two major TEEs currently not supported by \sysname are 
\qsee and \kinibi.
\qsee TAs wait for requests in a loop in their \texttt{entry} function.
Upon request reception, the TA handles the request in its 
\texttt{tz\_app\_command\_handler} function. 
Implementing this dispatching logic in \sysname requires loading and running the TA
 until it enters the waiting loop. Then, \sysname can set up 
the input and output buffers and call the command-handling function.
\qsee TAs are dynamically linked and the TEE-specific APIs 
all start with \texttt{qsee\_}. Extending \sysname with 
the most used \qsee APIs should be straightforward using greedy HLE.

\kinibi TAs similarly run from their entry point until a main loop, 
waiting for incoming requests. These requests are then handled in \texttt{tlApiWaitNotification}.
Similar to \qsee, \sysname needs to load the \kinibi TAs and run them 
until the main loop, then invoke the command handling function.
Unlike \qsee, \kinibi TAs are statically compiled and thus identifying the 
APIs requires a one-time effort as described in \autoref{sec:implementation}.

\textbf{Manual Effort to Support TEE-specific APIs.}
The reverse engineering and implementation effort to support a new TEE-specific 
API depends on the nature of the API. For some TEE-specific APIs, only the TA's code 
is required to emulate the API. These include 
API invocations that don't modify any state in the TA, such as logging functionality or 
API invocations in which the TA only inspects the return code. For such 
cases, a simple stub allowing the TA to continue execution is sufficient.
For more complex API invocations, the analyst has to reverse engineer the implementation 
of the API function. This may involve reverse engineering 
the library, the kernel, or other TAs, in the case of 
inter-process communication (IPC).

\textbf{Emulating TA to TA IPC.}
One class of TEE-specific APIs is wrappers around TA to TA IPC.
In its current implementation, \sysname implements this TA to TA IPC by hooking the GP
API invocations responsible for IPC, inspecting the destination TA, and returning the expected data. 
Future work could extend \sysname to connect multiple instances of \sysname to transparently handle IPC.

\textbf{Visibility of Library Code.}
A significant limitation of API-level interception is the fact 
that library code is not emulated. There could be bugs in the libraries 
used by the TAs, or the TA could be misusing the library.
It is also a limitation for analyzing the exploitability of a heap-related bug, 
since the TEE allocator's internals are relevant. 
Extending \sysname to load TEE libraries along with the TA is 
possible. As long as the loaded libraries use the same APIs as the TA, no additional 
work is needed. However, for libraries designed to interact directly with the TOS, 
\sysname needs to be extended to handle these TOS interactions.

\textbf{TA Harnessing.}
Fuzzing is one of \sysname's main use cases. For our fuzzing evaluation, we 
implemented harnesses for \numfuzztas TAs. These \numfuzztas TAs expect  
a straightforward input format with a single input buffer. This translates very easily 
to fuzzing, as the fuzzing input can simply be copied to the input buffer. 
The remaining TAs use a more complicated input format, with the TA expecting the input 
in multiple input buffers or as input values. Furthermore, many TAs require a specific state to trigger 
relevant functionality. Automatically generating harnesses for more involved 
input formats or to build up state is an open research problem; 
\sysname provides a platform to evaluate automatic harness generation for TAs.

\section{Related Work}\label{sec:relatedwork}

Both academia and industry demonstrated the pervasiveness and criticality of TA
vulnerabilities~\cite{beniaini2017trustissues,berard2018kinibi,busch2020finding,rollback,globalconfusion,Komaromy2021UnboxYourPhone,Li2024DiveIntoAndroidTA,AdamskiGuilbonPeterlin2019SamsungTrustZone,Makkaveev2019TrustZoneFuzzing,horta,trustdies,unearthing,AdamTC,zhao2021wideshears}.
Motivated by these findings, the research community proposed various analysis
techniques for TAs.

The most popular approach is dynamic analysis, as it enables fuzzing. 
PartEmu~\cite{partemu} executes TAs by partially emulating hardware and
software dependencies of the TOS. This powerful approach enables dynamic
analysis for TAs, but comes with questionable feasibility in real-world
scenarios where datasheets for hardware and source code for closed-source
components are missing.
SyncEmu~\cite{syncemu} manually rehosts a single TOS (Huawei's TrustedCore) 
and fuzzes TAs by using a NWCA-in-the-loop approach, forwarding TA requests made 
by the COTS device to the rehosted TA.

As full TOS emulation is infeasible in real-world situations, the focus shifts
to emulating TAs.  LightEMU~\cite{lightemu} emulates TAs at the system call
layer. LightEmu reports successfully emulating eight TAs across four TOSes, but
lacks details concerning the reverse engineering overhead of implementing the
system call stubs for the various TOSes.  Fan et al.~\cite{fan2025qsee} propose
an approach similar to \sysname---hooking at the API level to emulate QSEE TAs.
They emulate one QSEE TA from the Pixel 4. Their work focuses on the internals
of the Widevine TA to reproduce an n-day bug with their emulator and does not
discuss the feasibility of API-level hooking.  A number of ad-hoc industry
projects leveraged syscall or API-level hooking to emulate one or multiple
TAs~\cite{qseecyberes,Thalium2023PivotingSecureWorld,Komaromy2021UnboxYourPhone,AdamskiGuilbonPeterlin2019SamsungTrustZone,Li2024DiveIntoAndroidTA}.

Another research thrust for dynamic analysis of TAs investigates the feasibility of 
on-device analysis.
Crowbar~\cite{crowbar} presents an approach to TA fuzzing on a development 
device using ARM Coresight to extract coverage. This approach is limited by an 
analyst's ability to obtain such debugging capabilities (e.g., Coresight access). 
Similarly, DTA~\cite{dta} executes TAs in the normal world user space and forwards system calls to a custom proxy TA running in the secure world. Prior to DTA, Makkaveev~\cite{Makkaveev2019TrustZoneFuzzing} demonstrated this 
approach to fuzz QSEE TAs on the Google Nexus 6 device.
This approach requires a vulnerability to compromise the TEE to allow 
loading arbitrary TAs. 
TEEzz~\cite{teezz} fuzzes TAs without coverage feedback on-device, leveraging dynamic 
analysis on NWCAs to generate high-quality seeds. This approach to high-quality 
seed generation is complementary to \sysname's fuzzing mode.

An alternative to dynamic analysis of TAs is static analysis. Busch et al. use 
static analysis to uncover both rollback~\cite{rollback} and parameter type 
confusion vulnerabilites~\cite{globalconfusion} in TAs.

\sysname leverages the economies of scale for manual rehosting effort unlocked by 
the convergence of the TA ecosystem towards a common API---the GP TEE Internal Core API. 
In contrast to previous approaches to dynamic analysis of TAs, \sysname unites a promising 
direction for generic and cross-vendor TA emulation by providing a generic TA emulator in 
combination with a systematic approach to handle TEE-specific APIs that are not yet part of the emulator.

\section{Conclusion}\label{sec:conclusion}

\sysname is the first principled HLE approach for TAs. 
\sysname leverages the well-defined abstraction interface provided by the GP and 
libc APIs, which allows \sysname to emulate 39\% of all potentially reachable basic blocks. 
To handle TEE-specific functionality, we introduce greedy high-level emulation (HLE), a technique to prioritize 
manual rehosting effort. By supporting standard GP and libc APIs, as well as adding support for ten  
TEE-specific APIs via greedy HLE, 90\% of all basic blocks are reachable in \sysname. 
Most of these ten TEE-specific APIs can feasibly be replaced by GP/libc APIs and 
are thus straightforward to implement.

\sysname is able to emulate \numtasall{} TAs from four different TEEs. 
We demonstrate \sysname's practicality by using it to fuzz TAs 
and uncover \numvulns 0-day vulnerabilities, which we have responsibly disclosed 
to the affected vendors. \sysname's source code and artifacts are publicly available.

\def\UrlBreaks{\do\/\do-\do_}
\bibliographystyle{plain}
\bibliography{bibliography}

\end{document}